\documentclass[12pt,preprint]{aastex}
\begin{document}

\parindent=1.0cm

\title 
{THE STELLAR ARCHEOLOGY OF THE M33 DISK: RECENT STAR-FORMING HISTORY 
AND CONSTRAINTS ON THE TIMING OF AN INTERACTION WITH M31
\altaffilmark {1} \altaffilmark {2}}

\author{T. J. Davidge, \& T. H. Puzia \altaffilmark{3}}

\affil{Herzberg Institute of Astrophysics,
\\National Research Council of Canada, 5071 West Saanich Road,
\\Victoria, BC Canada V9E 2E7\\ {\it email: tim.davidge@nrc.ca, 
tpuzia@gmail.com}}

\altaffiltext{1}{Based on observations obtained with MegaPrime/MegaCam, a joint
project of CFHT and CEA/DAPNIA, at the Canada-France-Hawaii Telescope (CFHT)
which is operated by the National Research Council (NRC) of Canada, the
Institut National des Science de l'Univers of the Centre National de la
Recherche Scientifique (CNRS) of France, and the University of Hawaii.}

\altaffiltext{2}{This research used the facilities of the Canadian Astronomical 
Data Center, operated by the National Research Council of Canada with the support 
of the Canadian Space Agency.}

\altaffiltext{3}{Present Address: Department of Astronomy and Astrophysics,
Pontifica Universidad Catolica, Santiago, Chile 7820436}

\begin{abstract}

	Images recorded with MegaCam are used to investigate the recent ($t 
\leq 0.25$ Gyr) star-forming history (SFH) of the Local Group Sc galaxy 
M33. The data sample the entire star-forming disk, as well as areas 
immediately to the north and south of the galaxy. The properties of the stellar 
disk change near R$_{GC} = 8$ kpc. Within this radius the luminosity 
function of main sequence stars indicates that the star formation rate (SFR) 
has been constant with time during at least the past 250 Myr, while at larger 
radii the SFR has declined during this same time period. That the recent SFR in 
the inner disk has been constant suggests that M33 has evolved in isolation 
for at least the past $\sim 0.5$ Gyr, thereby providing a constraint on the 
timing of any recent interaction with M31. The color of the main 
sequence ridgeline changes with radius, suggesting a gradient 
in extinction of size $\Delta A_V/\Delta R_{GC} = -0.05$ magnitudes kpc$^{-1}$. 
The fractional contribution that young stars make to the total mass of the 
stellar disk changes with radius, peaking near 8 kpc. Evidence is also 
presented of systematic spatial variations in the SFH of the disk, such that 
the SFR during the past 100 Myr in the southern half of the galaxy 
has been $\sim 0.4$ dex higher than in the northern half. Finally, 
structures with sizes spanning many kpc that contain blue objects -- 
presumably main sequence stars that formed during intermediate epochs -- 
are identified near the disk boundary. It is argued that these are 
tidal features that were pulled from the main body of M33 and -- in some cases 
-- are the fossil remnants of star formation that occured in an extended disk 
during intermediate epochs.

\end{abstract}

\keywords{galaxies: evolution --- galaxies: spiral --- galaxies: individual(M33)}

\section{INTRODUCTON}

	The largest members of the Local Group have not evolved passively, but 
have experienced interactions that are consistent with hierarchal assembly. 
Perhaps the most overt signatures of this activity in the Local Group are the 
stellar streams that have been detected in the outer regions of M31 (e.g. Ibata et 
al. 2007; Tanaka et al. 2010). The location and morphology of 
the Sagittarious dwarf galaxy (Ibata et al. 1994), coupled with the presence 
of coherent streams in the Galactic disk (e.g. Helmi et al. 2006) and 
globular clusters that follow an age-metallicity relation that is distinct 
from that defined by the majority of Galactic globular 
clusters (e.g. Forbes \& Bridges 2010), indicate 
that the Galaxy also has not evolved in isolation. There is evidence that the 
effects of recent interactions on the properties of the Galactic disk have not 
been as great as they have been on the M31 disk (e.g. Hammer et al. 2007).

	As the nearest Sc galaxy, M33 is on a par with the LMC in serving as a 
benchmark for understanding the stellar content and evolution of late-type 
galaxies. However, during the past few years, evidence has been presented that M33 
and M31 may have interacted during cosmologically recent epochs, and such an 
interaction may have affected the properties of the M33 disk. Braun \& Thilker 
(2004) find an HI structure that appears to link M33 and M31, and Bekki (2008) 
models this as a tidal bridge that formed during an interaction 4 -- 8 Gyr 
in the past. Putman et al. (2009) investigate the possible orbital trajectories 
of M31 and M33 as constrained by their present-day locations and 
radial velocities. They conclude that at some point 
during the past $\sim 1 - 3$ Gyr the tidal radius of M33 may have been 
$\leq 15$ kpc due to an encounter with M31, which would have greatly affected 
the disk of M33. Simulations discussed by McConnachie et al. (2009) also 
support the notion of an interaction between M31 and M33 that may have disturbed 
the M33 disk. Their simulations suggest that the pericentric separation $\sim 
2.5$ Gyr in the past may have been as small as 40 kpc.

	The spatial distribution and kinematic properties of the 
interstellar medium (ISM) of M33 show signatures of a recent tidal interaction. 
The HI in the outer regions of M33 has an S-shaped morphology and is distributed 
over $28 \times 38$ kpc$^2$, with almost one fifth of the gas by mass located 
outside of the star-forming disk (Putman et al. 2009). 
While S-shaped structures are classical signatures of tidal 
interactions in simulations (e.g. Barnes \& Hernquist 1992), 
the arms of the HI emission associated with M33 are in the opposite direction 
of disk rotation, possibly suggesting an encounter in the opposite 
direction of M33 disk rotation. 

	The HI in M33 also has a high velocity dispersion 
when compared with other late-type galaxies, with 
a mean $\sigma = 18.5$ km sec$^{-1}$ (Putman et al. 2009). While high velocity 
dispersions in the outer regions of disks are not uncommon, and have been 
interpreted as the product of heating by halo sub-structures (e.g. Herrmann 
et al. 2009), the entire gas disk of M33 is kinematically hot. Block et al. 
(2007) argue that the warp in the ISM at 5 kpc could result from the accretion 
of gas with an angular momentum vector that was distinct from that of the main 
body of the disk, which might be expected to occur due to an interaction. 
Finally, while molecular clouds in M33 have some characteristics that are similar 
to their counterparts in the Milky-Way and M31 (Sheth et al. 2008), 
the ratio of atomic to molecular gas in M33 is high, and the 
fraction of molecular gas in giant molecular clouds is low when compared with 
the Galaxy (Verley et al. 2010). Both of these properties are consistent with 
a disturbed ISM.

\subsection{Stars and Gas in M33: A Brief Review}

	Stars and gas in M33 provide a fossil record of the events that 
shaped its past. Because of the obvious challenges presented by crowding, much 
of the previous work on the resolved stellar content of M33 has examined 
the outer regions of the disk. There are hints that the orbits of disk 
stars in M33 may have been disrupted, such that the radial age distribution 
differs from that in more isolated late-type spirals (Gogarten et al. 2010). 
In fact, it has been known for some time that stars with properties that are 
consistent with a disk origin appear at large radii in M33, 
as might be expected if the disk was disrupted. Davidge (2003) 
found that red giant branch (RGB) stars located between 14 and 17 kpc 
(7 -- 9 disk scale lengths) from the center of M33 have [M/H] $\sim -1 \pm 0.3$ 
dex. This metallicity is higher than expected for stars that belong to a 
classical metal-poor halo, but is within the range expected for disk objects. 
A similar situation occurs in M31, where moderately metal-rich stars are 
distributed throughout much of the outer regions of the galaxy (e.g. Bellazzini 
et al. 2003).

	Davidge (2003), Block et al. (2007) and Verley et al. (2009)
find evidence of a large population of C stars in the outer disk of M33. 
The discovery of a large number of C stars is suggestive of an 
elevated star formation rate (SFR) 1 -- 3 Gyr in the past, which is the likely 
time of an interaction with M31 (Putman et al. 2009). 
There are indications that these stars may belong to a population that is 
distinct from the inner disk population. Indeed, if -- 
as argued by Block et al (2007) -- the change that is seen in the $8\mu$m profile 
is driven by the circumstellar dust contents of M giants and C stars, then this is 
indicative of a change in stellar content. Block et al. (2007) 
further suggest that the formation of the C stars may be tied to the 
events that caused the warp in the M33 HI disk at R$_{GC} = 5$ kpc. 

	Barker \& Sarajedini (2008) explore the chemical evolution of the outer 
regions of M33. The chemical evolution deduced from CMD morphology does not 
follow that expected for a closed box; rather, a time-variable rate of infall 
is required to explain the enrichment history. Barker \& Sarajedini (2008) 
find that the peak rate of gas accretion occured $3 - 7$ Gyr in the 
past, during which time 50 -- 60\% of all gas that has fallen onto 
the disk was accreted. The rate of infall has since plunged, with only 10\% of all 
infalling gas being accreted during the past 3 Gyr. While Barker \& Sarajedini 
(2008) discuss these results in the context of protracted disk assembly, the 
drop in the infall rate some 3 Gyr in the past 
coincides with the timing of an interaction between M31 and M33 
as predicted by both Putman et al. (2009) and Bekki (2008). A modest 
infall rate at the present day would be expected if 
circumgalactic gas was stripped by an encounter with another galaxy. 

	Past studies have found evidence for systematic radial 
variations in the star-forming history (SFH) of the M33 disk. 
Bastian et al. (2007) find that the number of young stellar groups 
drops suddenly at radii $> 4$ kpc, which they attribute 
to a change in the star-forming ISM, and it is only in the central 
4 kpc that Park et al. (2009) find clusters with log(t) $\leq 7.8$.
The impact of systematic gradients in the SFH are also seen in integrated light. 
This is demonstrated by Munoz-Mateos et al. (2007), who 
find a radial gradient in the specific SFR (sSFR) of M33 based on 
measurements in the UV and IR.

	Cioni et al. (2008) examine the distribution of 
M giants and C stars in M33, and find that the C/M ratio 
peaks $\sim 35$ arcmin ($\sim 10$ kpc) from the galaxy center. The radial 
variation in C/M is likely a product of gradients in both mean 
metallicity and age, in the sense that younger mean ages and 
higher metallicities occur at smaller radii. Using C/M as a metallicity indicator, 
Cioni (2009) finds a progressive radial decrease in metallicity with increasing 
radius in M33 out to 8 kpc, while at larger radii the metallicity profile is flat. 
Cioni (2009) interprets the break in the metallicity 
profile as an interface between inner and outer disk/halo components.

	The spectral energy distribution of the M33 disk at visible 
wavelengths is consistent with insight-out formation (Li et al. 2004), although 
the radial age distribution of stars may be affected by processes other than 
the mechanics of disk assembly. Williams et al. (2009) use deep ACS images to 
examine stellar content over a range of radii in M33. They find that mean age 
drops as radius increases when R $< 8$ kpc, but 
at R $> 8$ kpc mean age increases with radius. 
Such a radial age inversion is consistent with simulations of disk 
evolution (e.g. Roskar et al. 2008; Sanchez-Blazquez et al. 2009), and can be 
understood in the context of inside-out disk formation coupled with 
the dynamical evolution of disk stars, which causes 
the outermost regions of the disk to be populated by stars that 
formed at smaller radii, but have migrated outwards. 
The most overt signature of such processes at work in M33 is the 
difference in scale lengths that characterize the distributions of young stars and 
old stars, in the sense that old stars are distributed over larger spatial 
scales than young stars (e.g. Verley et al. 2009; Davidge et al. 2011). 

	Metallicity gradients are a natural consequence of disk assembly in a 
hierarchal universe (e.g. Colavitti et al. 2009). However, 
if M31 passed close to M33 then tidal forces may have mixed gas throughout the 
disk, thereby flattening pre-existing gradients in the ISM. Rosolowsky \& Simon 
(2008) examine the emission spectra of HII regions that span a range of 
radii in M33 and find that $\Delta$[O/H]/$\Delta$R $\sim -0.027 \pm 0.012$. This 
is substantially smaller than the mean [O/H] gradient in the 39 disk galaxies 
studied by Zaritsky et al. (1994), for which $<\Delta$[O/H]/$\Delta$R$> = -0.058 
\pm 0.009$, where the uncertainty is the standard error in the mean.
A caveat is that giant HII regions, upon which many of the measurements 
in the Zaritsky et al. (1994) compilation are based,
may define steeper gradients than those computed from all HII 
regions, regardless of size (Magrini et al. 2010).

	Rosolowsky \& Simon (2008) find considerable 
scatter about the mean relation between [O/H] and radius in M33, 
which they attribute to local inhomogeneities in the ISM. An intriguing result 
is that HII regions with low [O/H] (and hence presumably low total 
metallicity) are found at small radii, suggesting that material that has 
undergone only modest amounts of chemical enrichment is present in the part 
of M33 that is expected to be the most chemically mature. 
The presence of such material could be due to the tidally-induced mixing of gas 
from the outer regions into the central regions of the galaxy. Alternatively, 
Magrini et al. (2010) argue that the comparatively low mean metallicity computed 
for HII regions in the central kpc of M33 may be a result of selection effects in 
surveys that require the detection of [OIII]4363, which is a diagnostic of 
electron temperature. Still, it is evident from their Figure 19 that 
the mean [O/H] in the central kpc remains lower than that at 1.5 kpc even if HII 
regions that do not have [OIII]4363 detections are included in the sample.

	Very young stars might be expected to track the metallicity trends 
defined by HII regions. However, blue supergiants (BSGs) in M33 define a steeper 
radial metallicity gradient than HII regions (U et al. 2009), with a slope 
that is consistent with the mean in the Zaritsky et al. (1994) HII sample. 
A possible explanation for this difference in gradients 
is that U et al. (2009) do not find stellar counterparts to 
the oxygen-poor HII regions detected at small radii by Rosolowsky \& Simon (2008); 
this may be a consequence of the modest number of BSGs studied.

	The ages of HII regions and PNe differ by at 
least a few hundred Myrs, and so insights into the recent chemical enrichment 
history of the M33 disk can be obtained by comparing abundance gradients 
deduced from these types of nebulae. Magrini et al. (2009) find that 
$\Delta$[O/H]/$\Delta$R = $-0.031 \pm 0.013$ amongst M33 PNe, and this agrees 
with the gradient measured from HII regions by Rosolowsky \& Simon (2008). 
Metallicity gradients are expected to steepen with time if there is infall, and 
Magrini et al. (2009) attribute the lack of evolution of the [O/H] gradient 
to a low accretion rate during the past few Gyr. Such a drop in infall rate during 
intermediate epochs is in broad agreement with the conclusion reached by 
Barker \& Sarajedini (2008) from an investigation of CMDs. Alternatively, 
a static abundance gradient with time could also signal the accretion of 
gas with low or modest amounts of enrichment, at a fortuitous rate that 
is sufficient to balance any increase in metallicity that may arise from 
the recycling of nuclear processed material into the ISM. 

	Some properties of M33 may not be typical of late-type spiral galaxies 
in general. For example, when compared with other late-type galaxies, 
the star-forming activity in M33 appears to be subdued. Indeed, while the 
sSFR of M33 is close to the midpoint of spiral galaxies in general, 
it is near the low end of galaxies of similar morphological type (Munoz-Mateos 
et al. 2007). The radial sSFR gradient in M33 is also steeper 
than in the majority of galaxies (Munoz-Mateos et al. 2007). This may 
indicate a low sSFR at small radii, although the circumnuclear stellar content 
of M33 is similar to that in other nearby late-type spirals (Davidge 2000; 
Davidge \& Courteau 2002).

	The gas within 8 kpc of the center of M33 appears to have 
a sub-critical density for triggering star formation (Martin \& Kennicutt 2001), 
and this is one possible explanation for the low sSFR when compared with other 
late-type spirals. Considering the low gas density, the chemical properties 
of M33 are suggestive of a star-forming efficiency (SFE) that is higher than 
that among other nearby galaxies, and the SFE increases to progressively 
larger radii (Magrini et al. 2010). That star-forming 
activity occurs on a large scale in an environment with a low density may 
be related to the shear rate (Martin \& Kennicutt 2001), which influences the 
frequency of collisions between molecular clouds, or an increase in ISM 
pressure caused by material falling into the galaxy, coupled with a kinematically 
hot interstellar environment (Putman et al 2009). The gravitational field of stars 
in the disk may also facilitate the collapse of star-forming material 
(e.g. Verley et al. 2010).

\subsection{The Current Study}

	The youngest stars in late-type galaxies play a key role in defining 
the classical morphological characteristics of these systems, and if M33 has 
experienced a recent interaction then residual signatures of such an event may be 
evident in the spatial distribution of these objects. Massive young stars are also 
brighter than the vast majority of older stars, and so are less prone to crowding 
in ground-based observations. In the present paper, $u*$ 
and $g'$ images obtained with MegaCam on the 3.6 metre Canada-France-Hawaii 
Telescope (CFHT) are used to investigate the properties of young stars in M33. 

	Two specific aspects of young stars in M33 are examined in this paper. 
First, we investigate the recent SFH of the M33 disk, with emphasis on how it has 
changed with time and location during the past few 100 Myr. There are 
considerable uncertainties associated with SFHs that are based on integrated 
light indicators, and these uncertainties are avoided by employing star counts as 
a probe of SFH. While area-to-area variations in SFR may occur within a 
galaxy due to the passage of spiral density waves or stochastic effects, 
when averaged over long timescales and large areas 
the SFRs of isolated star-forming disks are expected to be more-or-less 
constant with time. Aside from dense galaxy clusters, where there may be 
a high-density ambient intergalactic medium, galaxy-wide departures from a 
constant SFR in intermediate mass galaxies are likely triggered by 
tidal encounters. 

	Second, we search for young or intermediate age stars outside of the main 
body of the M33 disk. One might expect a mixed bag of such objects at large 
distances from the galaxy center. The outermost regions of the stellar disk may 
contain stars that migrated from smaller radii (e.g. Sellwood \& Binney 
2002), while the passage of spiral density waves might generate isolated areas 
of star formation (Bush et al. 2010). Tidal effects may pull gas from the inner 
regions of disks, and studies of interacting galaxies indicate that star formation 
is not uncommon in tidal debris fields (e.g. Neff et al. 2005). 
Tidal interactions can also re-distribute existing stars over large 
areas (e.g. Teyssier et al. 2009). If a tidal plume 
formed due to an interaction between M31 and M33 
then young/intermediate age stars, that either were present in the disk 
prior to the interaction or formed during the interaction, might be found 
in the circumgalactic environment.

	An important aspect of this study is the near-complete spatial 
coverage of the M33 disk. The only areas missed in the stellar disk 
are those that fall in the gaps between detector banks. While 
crowding is an issue within 2 kpc of the galaxy center, a large fraction of the 
brightest blue stars are still resolved there with these data (\S 6). 
Such complete areal coverage facilitates the suppression 
of stochastic variations in the SFH that may occur over kpc spatial 
scales. In this sense, the SFH deduced from our data is more akin to 
what might be derived from -- say -- fiber spectroscopic studies of distant 
galaxies, rather than from pencil-beam surveys of distinct 
areas within nearby galaxies that typically cover only a few kpc$^2$.

	Observations in $u*$ are another important aspect of this program.
The $u*$ filter is well-suited to investigations of 
intrinsically luminous main sequence stars, not only because they have 
effective temperatures that places the peak of their spectral energy distributions 
shortward of $0.4\mu$m, but also because of the enhanced contrast with respect 
to the (predominantly red) unresolved stellar body of M33 when compared 
with filters having longer central wavelengths. Observations in 
$u*$ are also beneficial when searching for diffuse ensembles of young and 
intermediate age stars in the outermost regions of the disk, 
where the fractional contamination from unresolved background galaxies, 
the majority of which have red colors (e.g. Davidge 2008; 
Mouhcine \& Ibata 2009), is substantial. An obvious downside to observing in $u*$ 
is that there is increased sensitivity to reddening variations 
when compared with filters that have longer effective wavelengths. 

	The distance modulus computed by Bonanos et al. (2006), $\mu_0 = 24.92$, 
is adopted for this study. This distance estimate is based on a 
geometric calibration (the orbital properties of stars in an eclipsing binary), 
and is in excellent agreement with recent distances computed from 
the properties of B supergiants and ACS-based observations of the RGB-tip 
(U et al. 2009). Galleti et al. (2004) compare various distance moduli 
of M33, and the results are summarized in their Figure 10. While the Bonanos et 
al. (2006) value is near the upper end of the distance estimates considered by 
Galleti et al. (2004), it still falls well within the scatter envelope. 
In any event, the basic results of this study will not change greatly 
if a distance modulus that is -- say -- 0.2 mag smaller were to be adopted.

	The line-of-sight extinction is assumed to 
originate in a uniform absorbing sheet in front of M33, and the 
adopted total extinction is the result of combining estimates of the 
foreground and internal components. The foreground component is taken from 
Schlegel et al. (1998), for which A$_B = 0.18$, while the internal component 
is that computed for M33 by Pierce \& Tully (1992), for which A$_B = 0.16$; thus, 
the total extinction is A$_B = 0.34$ magnitudes. In \S 4 it is 
demonstrated that this model correctly predicts the mean 
extinction towards main sequence stars in M33. Still, it should be kept 
in mind that dust in M33 is not uniformly distributed. U et 
al. (2009) find that line-of-sight dust extinctions among bright A and B 
supergiants vary by $\sim 0.16$ magnitude in $(B-V)$, with a mean E(B--V) = 
0.08. This mean is consistent with the combined foreground 
and internal reddening values used here. It is demonstrated in \S 4 
that reddening variations, as gauged by the width of the main sequence, 
are not substantial, with $\Delta E(B-V) \leq \pm 0.1 - 0.2$ throughout 
much of the M33 disk.

	The paper is structured as follows. Details of the observations 
and the photometric measurements are discussed in \S 2 and \S 3. The 
color-magnitude diagrams (CMDs) of stars and 
an assessment of reddening variations throughout the disk are presented in \S 4. 
The recent SFH throughout the disk is investigated in \S 5 by comparing 
the luminosity function (LF) of main sequence stars with models. 
The spatial distribution of young and intermediate age stars throughout the 
disk and its immediate surroundings is examined in \S 6. 
A summary and discussion of the results follows in \S 7.
 
\section{OBSERVATIONS \& REDUCTIONS}

	This study uses a mix of archival and newly recorded MegaCam 
images. MegaCam (Boulade et al. 2003) is the wide-field $0.35 - 1.1\mu$m imager 
on the 3.6 meter Canada-France-Hawaii Telescope (CFHT). The detector is a 
mosaic of thirty six $2048 \times 4612$ pixel E2V CCDs that are deployed 
in a $4 \times 9$ format. A single exposure covers 
$\sim 1 \times 1$ degree$^2$ with 0.18 arcsec pixel$^{-1}$ sampling.

	The locations of the five fields that are discussed in this paper 
are indicated in Figure 1. The data for the Center Field consists of archival 
$u*$ and $g'$ images that were recorded as parts of programs 2003BF15, 2004BH06, 
2004BH20, and 2004BF26. The majority of the exposures recorded for these programs 
have similar integration times, lunar phasing, and (sub-arcsec) image quality; 
hence, they form a homogeneous dataset.

	The archival data were downloaded from the CFHT archive 
hosted by the Canadian Astronomical Data Center (CADC), and 
were screened to remove exposures recorded during times of poor seeing, 
poor transparency, and/or bright lunar phase. A catalogue of the exposures 
that were retained for this program is provided in Table 1. Stars in 
the final combined images of the M33 Center Field have FWHM $\sim 0.9$ 
arcsec in both filters.

	The four other MegaCam pointings were recorded during 
semester 2009B, and these sample the northern 
and southern edges of the M33 disk. The fields overlap, allowing 
the field-to-field consistency of the photometric calibration to be 
checked, while also providing an empirical estimate of random errors in the 
photometry. Five 150 sec exposures were recorded per field 
in $g'$, and ten 440 sec exposures were recorded in $u*$. The data were recorded 
during grey time (i.e. partial lunar phase) to permit scheduling flexibility. 
A linear 5 point dither pattern was employed to facilitate the suppression of bad 
pixels and cosmic rays. Stars in the final combined images have FWHM $\sim 0.9$ 
arcsec in both filters.

	Initial processing of the data, which included bias 
removal and flat-fielding, was done with the CFHT ELIXER pipeline. To provide 
images of a tractable size for photometric measurements and to better monitor 
point spread function (PSF) variations across the wide MegaCam field, sub-mosaics 
of CCDs in $2 \times 3$ groups were constructed for subsequent processing -- 
each $4 \times 9$ MegaCam CCD mosaic was thus divided into $2 \times 3 = 6$ 
sub-mosaics. There are wavelength-dependent distortions that 
introduce offsets of a few pixels between the $u*$ and $g'$ 
images near the edge of the MegaCam field. These offsets were removed by using 
the GEOMAP/GEOTRAN tasks in IRAF to map the $u*$ images into the $g'$ 
reference frame. The final processing steps were to stack and combine the 
sub-mosiac images, and then trim the results to the area of common exposure time.

\section{PHOTOMETRIC MEASUREMENTS}

	The photometric measurements were made with the PSF-fitting routine 
ALLSTAR (Stetson \& Harris 1988). The source catalogues, PSFs, 
and aperture photometry that are used by ALLSTAR were 
obtained by running tasks in the DAOPHOT package (Stetson 1987). 
The PSFs for each sub-mosaic were constructed from 50 -- 100 
stars that were selected according to brightness, appearance, 
and an absence of bright neighbors. Faint stars in the vicinity of the PSF stars 
were removed in an iterative manner, being subtracted from the images using 
progressively improved versions of the PSF.

	ALLSTAR iteratively rejects sources that have poor 
photometric measurements. Despite such culling, a fraction of the sources in 
the photometric catalogues produced by ALLSTAR have 
measurements that are suspect, and these should be identified and rejected. As 
in previous photometric studies of stars in nearby galaxies (e.g. Davidge 2010), 
sources were culled from the photometric catalogue using criteria 
based on the uncertainty in the magnitudes calculated by DAOPHOT, $\epsilon$.

	All objects with $\epsilon > 0.3$ mag were excised from 
the catalogues, as useful photometric information can not be obtained for these 
sources, the vast majority of which are at the faint limit of the data. 
In addition, objects that depart from the trend between $\epsilon$ and 
magnitude defined by point sources in each field were also removed. 
As demonstrated in Figure 2 of Davidge (2010), the 
objects rejected in this step tend to be non-stellar in appearance, with 
the majority being galaxies, stellar blends, or cosmetic defects. At intermediate 
and brighter magnitudes this filtering may reject up to a third of all sources.

\subsection{Calibration}

	Photometric standard stars are observed during each MegaCam run, and 
the calibration information generated from these observations is inserted into the 
image headers as part of the ELIXER processing. The photometry used in this paper 
was calibrated using this information. The $u*$ measurements were transformed into 
$u'$ mganitudes. 

	Massey et al. (2006) investigated the UBVRI 
properties of the brightest stars in M33, and their catalogue can be used to 
check our calibration. The spatial coverage of the Massey et al. (2006) survey 
restricts these comparisons to the Central Field. The comparisons were restricted 
to only the brightest stars that fall outside of the main 
body of the disk to minimize the impact of crowding. The Massey et al. (2006) 
and MegaCam data were recorded during different epochs, and so variable stars 
will contribute to the scatter between the two sets of measurements.

	$u'$ and $g'$ magnitudes for stars in the Massey et al. (2006) 
catalogue were calculated using the empirical transformation relations found 
by Smith et al. (2002). The CMD of the stars that were used to check the 
measurements is shown in the top panel of Figure 2. While the sample 
contains sources with intermediate and blue $u'-g'$ colors, the majority of 
sources have red $u'-g'$ colors. This is a consequence of restricting the 
comparison sample to stars outside of the crowded star-forming disk of M33. 

	The difference between the two sets of measurements, $\Delta u'$, 
in the sense CFHT -- KPNO, is shown as a function of $u'-g'$ in the lower panel 
of Figure 2. The mean $\Delta u'$ is $0.04 \pm 0.06$ magnitudes, where the quoted 
uncertainty is the formal error of the mean. The Spearman correlation coefficient 
between $\Delta u'$ and $u'-g'$ is --0.34, indicating that 
there is not a significant correlation between these quantities.
The scatter about the calibrating relation between $u'-g'$ and $U-B$ in 
Figure 13 of Smith et al. (2002) is roughly $\pm 0.1$ magnitudes. If 
the four outliers with large $\mid \Delta u' \mid$, which are probably variable 
stars, are not considered then the majority of $\Delta u'$ values fall 
within the $\pm 0.1$ magnitude dispersion expected based on 
the scatter in the empirical relation between $u'-g'$ and $U-B$. 
We conclude that the good mean agreement between the CFHT and NOAO 
measurements indicates that the photometric calibration is reasonably sound.

\subsection{Artificial Star Experiments and Data Characterization}

	The uncertainty computed for magnitudes measured by ALLSTAR is based on 
the quality of the PSF fit, but does not account for systematic errors 
that might result from -- say -- crowding; thus, it gives only a lower 
limit to the actual uncertainties in the photometric measurements. Supplementary 
information about the photometric uncertainties in ensembles of 
stars, rather than individual sources, can be obtained from artificial star 
experiments. These experiments have the merit of also providing 
information about sample completeness.

	Main sequence stars are the primary targets of this 
investigation, and so the artificial stars were assigned magnitudes and 
colors that fall along the observed main sequence ridgeline. As 
with actual stars on the sky, artificial stars were only considered to 
be recovered if they were detected in both filters. The 
photometry files produced in the artificial star 
experiments were also culled using the $\epsilon$--based criteria 
described earlier.

	The incidence of blending and the random errors in the photometry 
both increase rapidly with magnitude when the completeness fraction drops below 
50\%, and so the magnitude at which 50\% completeness occurs provides a simple 
metric for gauging the faint end of the useable 
data range. The distribution of stars in the M33 disk means that 
the 50\% completeness magnitude changes with location in the 
Central Field, and the artificial star experiments indicate that 50\% completeness 
occurs near $u' = 24$ at R$_{GC} = 5$ kpc (18 arcmin), and $u' = 25.3$ at 
R$_{GC} = 9$ kpc (32 arcmin). For comparison, 50\% completeness occurs 
near $u' = 25.6$ throughout the fields that were observed in 2009B.

	The photometry was further characterized by comparing measurements made in 
overlapping sections of the fields. Such comparisons have the potential of 
revealing sources of error that are not addressed by the artificial star 
experiments. The difference in $u'$ between sources common to both the NE and NW 
fields was found to have a roughly constant dispersion when $u' < 24$, 
amounting to $\sigma = \pm 0.03$ magnitudes. This is larger 
than predicted by the artificial star experiments, which predict a dispersion 
of $\pm 0.02$ magnitudes at $u' = 22$.

	Artificial star experiments require an input PSF, and for ground-based 
observations the PSF typically is constructed from 
the data. One consequence is that the uncertainties deduced 
from artificial star experiments do not account for PSF-matching 
errors, which will be present at some level in all datasets, as it is not possible 
to characterize the PSF in a perfect way. The NE and NW fields 
overlap at the edges of the MegaCam science fields where the PSF can vary 
over small spatial scales, and the constant dispersion in the difference between 
the photometric measurements in these fields over a range of magnitudes 
is consistent with the behaviour expected from PSF-matching errors. 

	PSF-matching errors due to spatial changes in the PSF can 
be quantified by computing aperture corrections, 
which are the differences between magnitudes measured using large apertures and 
those determined from PSF-fitting, at various points across a field. 
The aperture correction will vary across the field if there are changes in the PSF 
that have not been tracked. We have computed aperture corrections for bright, 
isolated sources in the section of the NE field that overlaps with the NW field. 
The aperture correction varies systematically with location across the field, 
in the sense of increasing in size towards the edge of the MegaCam field. The 
amplitude of the variation of the aperture correction is 0.06 magnitudes, which 
is in excellent agreement with the residual dispersion discussed above. 
We thus conclude that PSF-matching errors introduce uncertainties of a few 
hundredths of a magnitude to these data.

\section{AN OVERVIEW OF THE STELLAR CONTENT}

\subsection{Star-Forming Complexes}

	The disk of M33 contains a number of young stellar complexes, 
and the photometry of stars in such systems provides an empirical means of 
assessing the impact of crowding on the photometry in general. To this end, 
four associations/star-forming complexes -- IC 133, A 31, A 116, and A 
127 -- were selected for study. The last three objects are from the catalogue of 
Humphreys \& Sandage (1980). It should be emphasized that these are not star 
clusters in the traditional sense, but giant star-forming complexes 
of the type discussed by Efremov (1995). The sizes of these complexes 
may reflect the spatial scales for star formation defined by 
turbulence (Elmegreen et al. 2003). 

	These four particular complexes were selected  
because they are in isolated environments, thereby 
making their boundaries relatively easy to 
identify when compared with similar complexes in denser portions of the disk. 
The geometric center and radius for each complex were 
estimated by eye, and the CMDs of objects within the extraction regions are 
shown in Figure 3. These complexes span a range of source densities. 
Aperture photometry indicates that IC 113, A 31 and A 116 have comparable 
surface brightnesses, whereas A 127 is roughly 1 mag 
arcsec$^2$ brighter in $u'$ and $g'$.

	Even though these complexes are in isolated portions of M33, there 
is non-negligible contamination from disk stars. To allow 
the extent of this contamination to be assessed, the CMDs 
of sources in an annulus surrounding each complex, that subtends the same total 
area on the sky as the cluster that it abuts, were also constructed, and 
the results are shown in Figure 3. The number densities of sources near 
the faint ends of the background CMDs are higher than in the CMDs of the 
adjacent clusters, demonstrating the diminished impact of crowding in the 
(comparatively) low density circumcluster environments.

	The main sequence is the dominant feature in the cluster CMDs. Because 
young stellar regions do not have hard boundaries, some fraction of the stars in 
the background regions are almost certainly cluster members; still, 
the brightest main sequence stars occur in the 
young complexes, rather than the background fields. 
Sources to the right of the main sequence are BSGs.

	The CMDs of the young stellar complexes are compared with 
isochrones from Girardi et al. (2004) in Figure 4. The models have Z = 0.008, 
which is close to the metallicity measured for BSGs in M33 by U et al. (2009). 
There is reasonable agreement between the predicted and observed main sequence 
color of A 31 over a large range of $u'$ magnitude. This suggests that the 
baseline reddening is appropriate for this complex, and 
that crowding does not appear to affect the photometric measurements.

	In the case of A 116, the stars with $u' < 
22$ are well matched by the 10 Myr isochrone, but at fainter magnitudes 
many sources fall redward of the predicted main sequence. 
Because A 116 is a star-forming region then it is possible that the red 
objects may be stars that have not yet collapsed onto the main sequence. 
Still, for these to be pre-main sequence stars then 
the evolutionary tracks of Cignoni et al. (2009) 
suggest that there must have been a significant star-forming episode 
in A 116 within only the past few Myr. For comparison, the main sequence 
turn-off of A 116 is well matched by the 10 Myr isochrone.

	The agreement between the observations and models is not as good for the 
other two complexes, with the locus of blue stars falling $\sim 0.1 - 0.2$ 
magnitudes redward of the zero age main sequence (ZAMS) for a young 
population. This could indicate that A$_V$ is $\sim 0.2 - 0.3$ magnitude 
higher for these complexes than the baseline value. 
The effect of adopting a higher line-of-sight extinction is demonstrated in 
Figure 4, where the dashed line shows the 10 Myr isochrone with 
A$_V = 0.2$ mag additional extinction. 

	While the agreement with the observed color of the main sequence 
is improved somewhat with a higher extinction, the agreement at 
fainter magnitudes remains poor, in the sense that the 
majority of stars still fall to the right of the predicted ZAMS. 
A comparatively red main sequence near the faint end of 
IC 133 would occur if this structure contains stars that span a range 
of ages. Indeed, the main sequence in the CMD of this complex 
can be matched if it contains stars that formed during 
the past $\sim 100$ Myr. This timescale falls within the expected lifetime 
of large star-forming complexes (e.g. Davidge et al. 2011). 
However, this explanation does not hold for 
A 127, where the number of stars plunges near $u' = 22$. 

	In summary, even though these are relatively dense stellar aggregates, 
the photometry of stars with $u' < 22$ tends to be well matched by model 
isochrones. The one exception is A 127, which has the highest surface brightness 
of the four complexes studied. However, even in this complex the isochrones 
pass through the main sequence at $u' \sim 20$. 
These comparisons indicate that the photometry of main 
sequence stars with ages $\leq 10$ Myr should be reliable in 
relatively dense environments of M33, and in \S 6 this is validated by 
number counts of sources throughout the disk. Of course, in the lower 
density portions of the disk the faint limit of the data will be such 
that stars with ages that are well in excess of 10 Myr can be examined.

\subsection{The M33 Disk}

	The $(u', u'-g')$ CMDs of stars in various 
radial intervals in the Central Field are shown in Figure 5. 
The distances given in each panel are measured from the center of M33 in 
the plane of the disk. A disk inclination of 54 degrees was 
assumed (Pierce \& Tully 1992).

	The influence of individual large star-forming 
complexes are suppressed when stars from large swaths of the M33 disk are 
combined. The main sequence forms a prominent plume with $u'-g'$ between 0.0 
and 0.5 in the R$_{GC} < 10$ kpc CMDs, and the diffuse spray of objects to the 
immediate right of the main sequence is dominated by BSGs. 
Source confusion increases with stellar density, and this 
can clearly be seen in the comparatively shallow faint limit in the CMDs for 
R$_{GC} < 4$ kpc. At the bright end, crowding is mitigated to some extent by 
the blue wavelength coverage of the $u*$ filter, which avoids the peak in the 
spectral-energy distribution of the majority of disk stars. 
The brightest blue objects in Figure 5 have $u' \sim 17$ when R$_{GC} 
< 8$ kpc, and that this peak brightness does not change with 
radius when R$_{GC} < 8$ kpc suggests that the incidence of blending is not 
significant amongst the brightest sources.

	The properties of the CMDs in Figure 5 might be expected 
to change with radius based on results presented in previous 
studies. Munoz-Mateos et al. (2007) and Verley et 
al. (2009) present evidence that the fractional contribution made by young 
stars to the overall stellar content increases with radius in M33, although this 
trend ultimately reverses in the outer galaxy (Williams et al. 2009). 
Bastian et al. (2007) and Park et al. (2009) find that the number of young 
stellar groups and clusters drops near R$_{GC} = 4$ kpc. In contrast, 
Thilker et al. (2005) find that the FUV--NUV color is constant 
across the disk, with a value that is consistent with a flux-weighted age in the 
range 200 -- 300 Myr, suggesting that the young stellar mixture does not change 
with radius. In fact, there is no obvious change in the properties of the 
main sequence stars and BSGs in the R$_{GC} < 8$ kpc CMDs in Figure 5.

	Thilker et al. (2005) find that UV emission from the M33 disk drops 
at a radius of 30 arcmin (R$_{GC} = 8.5$ kpc), signalling a rapid 
decrease in recent star-forming activity. Signatures 
of such a drop in star-forming activity should also be 
evident in the $(u', u'-g')$ CMDs. In fact, a marked change 
in the brightness of the MSTO between the 6 -- 8 kpc and 8 -- 10 kpc CMDs is 
evident in Figure 5. 

	The mean color of main sequence stars in a given magnitude interval 
changes with radius. This is demonstrated in Figure 6, where the histogram 
distributions of $u'-g'$ colors for sources with $u'$ between 20.5 and 21.5 in 
four annuli are compared. The prominent bump in the color distributions is 
due to main sequence stars, and the peak of this feature 
is displaced to progressively redder values with decreasing 
distance from the center of the galaxy. A least squares fit to the 
peaks of the color distributions indicates that 
$\Delta(u'-g')/\Delta(R_{GC}) = -0.021 \pm 0.004$ kpc$^{-1}$. The recent SFH of 
M33 does not change with radius in this part of the galaxy (\S 5), and so the 
change in main sequence color is not due to a systematic gradient in 
the SFH. Rather, the radial trend in mean color could be attributed 
to systematic changes in the amount of extinction. Adopting the 
reddening law in Table 6 of Schlegel et al. (1998) then gradients 
amounting to $\Delta E(B-V)/\Delta R_{GC} = 
0.015 \pm 0.003$ kpc$^{-1}$ and $\Delta A_V/\Delta R_{GC} = -0.05 \pm 0.01$ 
kpc$^{-1}$ could explain the trends in peak main sequence color. 
Davidge (2007) also found evidence for higher levels of extinction 
among the youngest stars in the central few kpc of the Sc galaxy NGC 2403.

	The widths of the color distributions do not change with radius 
when R$_{GC} < 6$ kpc. The artificial star experiments predict a dispersion 
in $u'-g'$ of $\pm 0.04$ mag in the 4 -- 6 kpc annulus, whereas the observed 
dispersion is $\sigma_{u'-g'} = \pm 0.135$ mag. The width of the main sequence 
in these observations is thus not dominated by observational errors. A number of 
factors, such as evolution away from the ZAMS, 
differential reddening, binarity, and photometric variability will 
contribute to a main sequence that is wider than 
expected from observational uncertainties alone. An upper 
limit to the amount of differential reddening can be obtained by 
assuming that it is the dominant source of dispersion in $u'-g'$. 
After subtracting in quadrature the contribution made by observational 
errors, then we find that $\Delta E(B-V) \leq \pm 0.10$ and $\Delta A_V \leq 
\pm 0.3$ throughout the inner disk. For comparison, the scatter in reddening 
found by U et al. (2009) from AB supergiants is $\pm 0.04$ mag in E(B--V).

	The CMDs are compared with Z = 0.008 isochrones from Girardi 
et al. (2004) in Figure 7. The isochrones track the 
locus of main sequence stars, as expected if the baseline reddening 
model applies to the majority of young stars throughout the 
M33 disk. Comparisons with the isochrones further indicate that the youngest stars 
in the 4 -- 6 kpc interval have ages $\leq 10$ Myr. However, at larger radii the 
ages of the youngest stars increases with galactocentric distance, 
such that in the 12 - 14 kpc interval the youngest stars have ages $\sim 100$ Myr.

	Star counts provide insights into the recent SFR. We consider 
the statistics of stars with ages $\leq 10$ Myr -- which have masses $< 20$ 
M$_{\odot}$ -- as these objects are present throughout the central 8 kpc of 
M33 (\S 6) and are less prone to crowding-related issues than fainter objects.
With a global SFR of 0.7 M$_{\odot}$ year$^{-1}$ (Blitz \& Rosolowsky 2006), then 
$7 \times 10^6$ M$_{\odot}$ of stars will have formed in the past 10 Myr 
throughout M33. Adopting the Kroupa (2001) initial mass function (IMF) from 
0.08 to 100 M$_{\odot}$, then $8 \times 10^5$ M$_{\odot}$ of stars with masses 
in excess of 20 M$_{\odot}$ will have formed. This translates 
into $\sim 10^4$ main sequence stars, and such objects will have 
$u' \leq 19.5$. For comparison, $\sim 3 \times 10^3$ objects with $u' \leq 19.5$ 
are seen throughout M33. Some of these bright objects 
are undoubtedly binaries or unresolved compact star clusters, and so 
the actual number of massive main sequence stars will be less than this.

	There is roughly a factor of $\sim 3$ difference between 
the observed number of very bright stars and the number expected if the 
SFR $= 0.7$ M$_{\odot}$ year$^{-1}$. Possible explanations 
for this discrepancy are that (1) the SFR has been lower than 0.7 M$_{\odot}$ 
year$^{-1}$ during the past 10 Myr, (2) many very young stars are 
missing, either because they are in unresolved blends or are 
heavily obscured, and hence are not detected, and/or (3) that the IMF in M33 may 
be skewed to the production of fewer massive stars than predicted by Kroupa 
(2001). We consider the first two of these to be more likely than the third, given 
that model LFs constructed with a Kroupa (2001) IMF match the slope of 
the observed LF of bright main sequence stars throughout M33 (\S 5). In \S 7 we 
argue that a SFR of 0.7 M$_{\odot}$ year$^{-1}$ is probably not sustainable in 
M33 for a prolonged period of time given the mass of the ISM.

\subsection{The Outer Disk}

	The $(u', u'-g')$ CMDs of sources in the NE, NW, SE, and SW fields are 
shown in Figure 8. The source densities in these fields are lower 
than in the Center Field, and so the CMDs in Figure 8 extend to fainter 
magnitudes than is typical of the Center Field. The majority of sources 
with $u'-g' \geq 1$ and $u' > 24$ are background galaxies, although 
core Helium burning stars will also occupy this part of the CMD 
if an intermediate age population is present. Blue HB stars belonging to an 
old metal-poor component are too faint to be detected with these data. 

	There is a well-populated main sequence 
in the CMDs of the NE and NW fields, and many of the brightest 
main sequence stars belong to the two star-forming complexes in 
the northern spiral arm of M33 that were studied by Davidge et al. (2011). 
Bright main sequence stars are also seen in the CMD of the SW field, although 
there are fewer stars than in the NE and NW fields. The CMDs of the portions 
of the NE and SW fields that have the highest stellar densities 
are compared with Z = 0.008 isochrones from 
Girardi et al. (2004) in Figure 9. The areas from which sources 
were extracted to construct the CMDs in Figure 9 are indicated in Figure 17. 

	Whereas the brightest main sequence stars near the northern edge of the 
disk follow the 10 Myr isochrone, the majority of stars in the southern 
part of the disk have ages $\geq 10$ Myr. The red envelope 
of the main sequence in both CMDs tracks the 
redward extent of main sequence evolution predicted by the isochrones. This 
is consistent with the majority of blue sources with $u' \geq 24$ in these areas 
being intermediate age stars, rather than background objects.

	A rich population of objects with $u'-g' \sim 0.4$ and $u' > 
23.5$ are seen in the CMDs of all four fields. The photometric 
properties of these objects are consistent with them 
being main sequence stars with ages $\geq 100$ Myr. 
While in reality a substantial fraction are either blue 
galactic nuclei or star-forming regions in background galaxies, 
it is demonstrated in \S 6 that some of these faint blue objects define 
structures in the outer regions of the M33 disk, and so are probably stars that 
belong to M33. 

\section{THE LFs OF MAIN SEQUENCE STARS AND THE STAR-FORMING HISTORY}

\subsection{Model LFs}

	The faint limit of these data is sufficient to allow the SFH 
during the past few hundred Myr to be probed using the $u'$ LF of main 
sequence stars. The structural properties of main sequence stars 
are subject to fewer uncertainties than those of more evolved objects, making 
model LFs of main sequence stars potentially more reliable probes of SFH than the 
LFs of stars in advanced stages of evolution. The affect of uncertainties in 
the advanced stages of stellar evolution on SFHs has been investigated by 
Melbourne et al. (2010), who compare SFHs of the 
dwarf galaxy KKH 98 deduced from AGB and main sequence stars. These two stellar 
types yield SFHs that differ significantly during intermediate epochs, and this is 
attributed to uncertainties in the physics of AGB evolution. Still, our 
knowledge of main sequence evolution is not ironclad, and uncertainties in the 
mechanics of mass loss, coupled with an incomplete knowledge of convection -- 
the dominant source of energy transport in the central regions 
of massive stars -- are some of the deficiencies that affect 
models of massive stars as they evolve off the ZAMS, 
but are still burning Hydrogen in their core.

	Binarity is not an issue when modelling more evolved stages of 
evolution at visible wavelengths, as the more massive star in a pair evolves 
from the main sequence first. The light of the primary then 
swamps that of its less massive (and hence less evolved) companion(s). 
Binarity is a concern when modelling the characteristics of main sequence stars at 
visible wavelengths as brightness differences between the primary stars and 
any companions are much smaller than if one star is highly evolved. 
However, as long as the binary fraction and mass ratio characteristics do not 
change with the mass of the primary then the overall {\it shape} of main sequence 
model LFs will not be affected greatly by binarity. In recognition of this 
issue, comparisons with model LFs are restricted to LF shape, rather than the 
absolute predictions of star numbers predicted by the models.

	Model LFs were constructed from isochrones in the Girardi 
et al. (2004) compilation using routines in STARFISH (Harris 
\& Zaritsky 2001). IMFs with mass function exponents 
$\alpha = -2.7$ (Kroupa et al. 1993) and $\alpha = -2.3$ (Salpeter 
1955) for massive stars were considered. The models assume Z = 0.008. There is a 
radial metallicity gradient in the inner disk of M33 (e.g. U et al. 
2009), and models with lower metallicities than this may 
be more appropriate for comparisons with outer disk LFs. However,
the structural properties of ZAMS stars in the range of stellar masses 
considered here ( $> 3$ M$_{\odot}$) is not sensitive to metallicity.

	The model LFs are used as interpretive 
tools, and no attempt is made to model particular features in the observations. 
To this end, models were generated for two SFHs that provide useful diagnostic 
information. One set of models assumes simple 
stellar populations (SSPs), in which the component stars are coeval 
and have a single metallicity. While SSP models are strictly appropriate only 
for star clusters, these models may approximate the bright end 
LFs of systems that have undergone a single large star-forming event 
during recent epochs.

	The second set of models assumes a SFR that is constant with time. Such a 
SFH might be expected to dominate in undisturbed disks if stochastic effects are 
suppressed by averaging over large areas and 
long time intervals. The metallicities of stars that form 
during constant SFR conditions will grow steadily with 
time, as previous stellar generations process and recycle interstellar material. 
While our models do not account for such enrichment, it is anticipated that 
this will not significantly affect the results. Indeed, the age-metallicity 
relation in the M33 disk during the time interval that we explore -- the past few 
hundred Myr  -- has been flat (Magrini et al. 2009). Even if 
an age-metallicity relation that is appropriate for the Solar neighborhood 
were to hold, it would result in only marginal enrichment over the 
$\sim 200$ Myr time interval that is sampled by the stars examined here.

\subsection{Comparisons with Model LFs}

	The LFs of objects in the M33 disk with $u'-g'$ between --0.5 and 0.5, 
which is the color interval that is dominated by massive main sequence 
stars, are shown in Figure 10. The faint limit in the 0 - 2 and 2 - 4 kpc 
LFs is markedly brighter than at larger radii owing to the higher stellar 
densities in these intervals. Source counts from the outer 
regions of the NE field, where the density of bright stars that 
belong to M33 should be negligible, have been subtracted 
to account statistically for contamination from foreground stars and background 
galaxies. This correction has only a modest impact on the R$_{GC} \leq 10$ kpc LFs.

	The shape of the LFs in Figure 10 changes with radius. Williams et al. 
(2009) conclude that the stellar content of M33 changes significantly near 8 kpc, 
and Thilker et al. (2005) find a drop in the level of UV emission and a 
change in the FUV--NUV color at this point. It is thus worth noting that the 
character of the LFs in Figure 10 also changes near R$_{GC} = 8$ kpc. The LFs of 
sources with R$_{GC} \leq 8$ kpc follow a single power-law with a common 
exponent, whereas at larger radii the LFs at the bright end flatten. The 
number of sources with $u' \leq 20$ also drops near R$_{GC} \sim 10$ kpc, 
although this may be due to small number statistics. 

	Three LFs from Figure 10 are compared with the SSP and 
constant SFR model LFs in Figures 11 and 12. The models have been scaled to match 
the number densities of objects with M$_{u'}$ between --2 and --4. It is apparent 
from Figure 11 that the SSP models are much flatter than the observed LFs, 
indicating that the young stars in the M33 disk formed over a range of epochs, 
and not in a single large-scale event.

	In contrast to the SSP models, the constant SFR models in Figure 12 are an 
excellent match to the 6 -- 8 kpc and 8 -- 10 kpc LFs in the interval M$_{u'} 
\leq -2$. The agreement is noteworthy given that no effort was made 
to tune the models to match the observations. The agreement with the 10 -- 12 
kpc LF is much poorer, in the sense that the slope 
of the observed LF is not reproduced at the faint 
end, and the fractional contribution made by older stars is higher 
than expected from a constant SFR in this radial interval. The absence of 
stars with M$_{u'} < -5$ in this radial interval may be due to small number 
statististics, as only a few are expected in this brightness range under the 
assumption of a constant SFR.

	The ISM of M33 is not distributed symmetrically throughout the galaxy 
(e.g. Verley et al. 2010), and so region-to-region variations in the SFH of the 
M33 disk might be expected. To assess the amplitude of possible large-scale 
azimuthal SFH variations throughout the M33 disk, the LFs of 
sources in two of the annuli examined in Figures 11 and 12 were 
constructed using 90 degree azimuthal gathers, as opposed to 
the 360 degree gather used to construct the LFs in Figure 10. 
The four resulting LFs in each radial interval are compared in Figure 13.

	There are only minor quadrant-to-quadrant variations at the faint end 
of the LFs in Figure 13. This is perhaps not unexpected given that 
progressively fainter magnitude intervals contain main sequence stars 
that formed over wider ranges of epochs than at the bright end, 
with the result that variations in SFH average out. However, the situation is 
very different near the bright end. While small number 
statistics are a factor amongst the brightest stars, 
in the interval between M$_{u'}= -3.5$ and --1.5 the quadrant-to-quadrant 
differences are larger than predicted by random errors alone. 
A systematic difference between the LFs of the northern and southern halfs of the 
galaxy is evident in this magnitude interval, in the sense that the source density 
in the south is $\sim 0.4$ dex higher than in the north. This suggests 
that the SFR during the past $\sim 100$ Myr in the southernmost part of the 
disk has been $\sim 2\times$ that in the northernmost part. This is consistent 
with the southern part of the disk having a higher HI concentration than the 
northern part (e.g. Verley et al. 2010).

\section{THE PROJECTED SPATIAL DISTRIBUTION OF MAIN SEQUENCE STARS}

\subsection{The Large-Scale Distribution of Disk Stars}

	The distribution of sources in the Central Field, selected 
from the $(u', u'-g')$ CMD using the photometric 
criteria defined in Figure 1 of Davidge et al. 2011 for 
characteristic ages of 10 and 100 Myr, are shown in Figure 14. 
The on-sky source distribution is shown in the top row, while the de-projected 
distribution, approximating the appearance of M33 as if viewed face-on, is shown 
in the bottom row. Depth effects have only a minor impact on the de-projected 
distribution given the moderate inclination of M33 to the observer.

	The star counts drop markedly at 25 -- 30 arcmin radius, or 7 -- 8 kpc, 
and this is the point at which UV emission also drops (Thilker et al. 2005).
It is also evident from Figure 14 that stars in the 10 and 100 Myr 
samples have very different distributions, in that much more 
clustering and spiral structure is evident in the 10 Myr sample than in the 
100 Myr sample. The lack of objects in the 100 Myr sample throughout the 
central regions of M33 is due to crowding.

	The uniform distribution of sources in the 100 Myr 
sample is largely a consequence of the random motions imparted to objects 
through interactions with giant molecular clouds, 
which in turn causes them to diffuse from their places of birth with time. 
Other processes, such as the churning of stellar orbits by spiral density
waves (Sellwood \& Binney  2002), further blur the stellar distribution, 
although these are expected to act over time spans of a few 
disk rotations, and so should not affect the locations of the stars that 
are considered in the present study, the oldest of which have an age of 
only $\sim$ one disk rotation.

\subsection{The Radial Distributions of Young and Old Stars}

	Bakos et al. (2008) find that the luminosity-weighted ages of 
Type II disks (i.e. those that -- like M33 -- have a downward-breaking 
light profile at large radii) at visible wavelengths is 
a minimum at the break radius. Based on the color profiles of these 
systems, they argue that the break in the light profile is a consequence of 
the luminosity-weighted age profile of the disk, rather than a discontinuity in 
the disk mass distribution. There is a possible physical driver for this 
behaviour, as cosmologically-based models of late-type 
spiral galaxies find that the ratio of gas to stellar mass peaks at the break 
radius, fostering a higher proportion of young stars at this point than elsewhere 
in the disk (Martinez-Serrano et al. 2009). 

	How do the radial distributions of young and old stars compare in M33? 
Verley et al. (2009) investigated the light profile of M33 at wavelengths ranging 
from $3.6\mu$m to $160\mu$m. While the light at mid and far-infrared wavelengths
is thermal in origin, and hence traces interstellar and circumstellar material, at 
$3.6\mu$m the light originates predominantly from the photospheres of cool stars, 
and so observations at this wavelength can be used as a proxy for overall 
stellar mass. With the exception of an upturn in the innermost regions of 
the galaxy, the $3.6\mu$m light profile of M33 shown in Figure 2 of 
Verley et al. (2009) follows a single power law between 1 and 9 kpc; thus, there 
is no evidence for a break in the stellar mass distribution near 8 kpc. 

	The number density of objects in the 10 Myr sample is compared with 
the $3.6\mu$m integrated light profile from Figure 2 of Verley et al. (2009) 
in Figure 15. The $3.6\mu$m relation has been 
adjusted to account for the slightly larger distance to M33 
adopted for the present study. The zeropoint of the $3.6\mu$m profile has 
also been adjusted so that the curve passes through the 
midpoint of the 10 Myr star counts. 

	The 10 Myr star counts define a trend that is flatter than the $3.6\mu$m 
integrated light profile at R$_{GC} < 8$ kpc, indicating that the 
contribution made by young stars to total stellar mass grows with increasing 
radius in the central 8 kpc. In addition, that the 
10 Myr star counts define a more-or-less linear relation 
for R$_{GC} \geq 1.5$ kpc suggests that incompleteness is not 
an issue for these very bright stars throughout much of the M33 disk. The 
number density of young stars drops near R$_{GC} \sim 8$ kpc, whereas the 
character of the $3.6\mu$m profile does not change at this point. 
The HI distribution of M33 has been mapped over a large area, and 
there is a localized maximum in HI emission near 7 kpc (e.g. 
Figure 3 of Corbelli 2003), which is close to the radius at which young stars 
make the largest contribution to the stellar mass density. The 
contribution that the youngest stars make to the total stellar mass drops 
near R$_{GC} = 8$ kpc; the radial distributions of young and old stars in the M33 
disk are thus consistent with that found in other systems by Bakos et al. (2008).

	In \S 5 it was shown that the recent SFH 
of the M33 disk changes near 8 kpc, in the sense that at smaller 
radii the SFR during the past few hundred Myr has been 
continuous, while outside the break radius there has 
been reduced star-forming activity over the past few tens of Myr when compared 
with the past few hundred Myr. The drop in the 
number density of very young stars near R$_{GC} = 8$ kpc, coupled with the 
continuous trend defined by $3.6\mu$m light, suggests that the SFR at large radii 
was not elevated during intermediate epochs. Rather, it is more likely that 
the SFR in the outer disk dropped during the past $\sim 100$ Myr.

	The decrease in the numbers of young stars at R$_{GC} > 8$ kpc is probably 
due in part to the asymmetric distribution of gas in the outer regions of M33. 
To the extent that HI traces molecular material, 
the HI distribution mapped by Putman et al. (2009) suggests that large portions 
of the circumdisk environment are devoid of star-forming material. 
Another factor is that the density of star-forming material throughout 
the M33 disk is sub-critical (Martin \& Kennicutt 
2001), and the probability of breeching the threshold for star 
formation is low in the outer regions of M33. Finally, the basic 
properties of star-forming material may change with 
radius in M33. Evidence for this comes from the physical 
sizes of molecular clouds in M33, which appear to decrease in size 
towards larger radii (e.g. Gratier 2010). Such a trend will affect the sizes of 
star-forming regions at large radii, which in turn may affect the chances of 
forming massive stars.

\subsection{The Spatial Distribution of Young Stars on Sub-Kiloparsec Scales}

	We use the star-star separation function (S3F) to investigate stellar 
grouping throughout the M33 disk. The S3F, which was defined by Davidge et al. 
(2011), is the histogram distribution of separations between all possible 
stellar pairings. Davidge et al. (2011) used the S3F to investigate 
the distribution of main sequence stars in the peripheral regions of M33, and 
significant projected spatial cohererence was detected among the youngest 
stars over separations $\leq 0.7$ kpc. The degree of 
coherence in the sample of objects studied by Davidge et al. (2011) was found to 
decrease with increasing stellar age, and it was concluded that star-forming 
complexes in this environment dissipate over time spans of a few tens of Myr.

	The S3Fs of M33 disk stars in two radial intervals are compared in Figure 
16. The current investigation is restricted to stars in the 10 
and 40 Myr samples because of the substantial level of incompleteness 
in the 100 Myr sample at small radii. In addition, as we are 
interested in stellar grouping on kpc and smaller spatial scales, 
the comparisons are limited to separations $\leq 2$ kpc, or roughly one 
disk scale length. The signal in each interval has been normalized to the number 
of objects with separations between 0.5 and 1.5 kpc (110 to 330 arcsec). 

	The shape of the S3Fs in Figure 16 differ from those examined by 
Davidge et al. (2011), in that prominent peaks at moderate separations 
are abscent. Davidge et al. (2011) studied a comparatively 
small area on the sky in the outer disk of the galaxy
that included two large star-forming complexes, so that there was 
a bias towards small separations. In contrast, the S3Fs in Figure 16 
are constructed from stars that cover large swaths of the disk, so that 
the influence of single star-forming complexes on the S3F are suppressed. 

	The S3Fs of the 40 Myr sample are
steeper than those in the 10 Myr sample at separations between 0 and 
200 arcsec (i.e. between 0 and 1 kpc). There is thus greater 
spatial coherence among stars in the 10 Myr sample with separations $< 60$ 
arcsec ($< 250$ parsecs) than in the 40 Myr sample. The reduced 
signal at small separations in the 40 Myr sample indicates that stars in the disk 
of M33 become less tightly grouped over timescales of only a few tens of Myr, 
due to the diffusion of stars from their places of birth.

\subsection{Diffuse Structures at the Disk Boundary}

	Structures of a tidal origin lack a dark matter cocoon, and so their 
morphology can evolve over timescales that are short when compared with 
classical galaxies. Indeed, while the dissruption timescale for structures 
in the outer regions of galaxies is comparatively long (e.g. Johnston 
et al. 1996), the prognosis for the long term survival of 
tidal structures is poor, with the vast majority expected to dissipate 
over time spans of only a few hundred Myr (Bournaud \& Duc 2006). 
Still, some structures may survive over Gyr or longer timescales 
if the interaction geometry is favorable; for example, long-lived 
tidal features may form if the orbit of the perturber is more-or-less 
coplanar with the disk of the host galaxy (Bournaud \& Duc 2006).

	The stellar content of a tidal feature provides a direct means of 
probing its history. A critical piece of information that can be gleaned 
from the stellar content is the age of the structure, which in turn indicates 
when the interaction occured. A tidal arm that is pulled from a star-forming 
disk in an interaction that occured during cosmologically-recent 
epochs should contain stars that span a wide range of ages, reflecting the 
extended SFH of the original disk. If the donor was actively forming stars before 
the interaction then some of the youngest stars in a tidal stream will have formed 
only shortly before the interaction. However, these may not be the youngest stars 
in the structure, as stars may form for some period of time after an interaction 
if local density enhancements persist in the gaseous component of the debris field.

	Evidence for low-level on-going star-forming activity is seen 
in extragalactic debris fields. The nearest example is the region between M81 and 
M82 (e.g. de Mello et al. 2008), even though the interaction occured 
a few hundred Myr in the past (Yun et al. 1994). In addition to star-forming 
structures, the M81--M82 debris field also contains a number 
of low density young stellar groupings that appear to have formed during the 
past few tens of Myr, but are not forming stars at present (e.g. Davidge 2008; 
Mouchine \& Ibata 2009). These objects suggest that there was likely larger scale 
star-forming activity in the debris field in the not too distant past. 

	Whether or not stars form depends on the conditions in the debris field, 
including the density of star-forming material. 
There is evidence of tidal material in the circumdisk environment of M33. 
The ridgeline of the HI distribution mapped by Putman et al. (2009) 
defines an S-shaped morphology, with arms that extend to the north west and 
to the south east of the disk. However, whereas the HI density 
throughout much of the M81--M82 debris field is $10^{21} - 10^{22}$ cm$^{-2}$ 
(Yun et al. 1994), comparable densities are found only in the main disk of M33 
(Putman et al. 2009). Indeed, the HI density in the arms that emerge to the north 
west and south east of the M33 disk is an order of magnitude lower than is seen in 
much of the M81--M82 debris field.

	A stellar structure that may coincide with one arm of the HI emission is 
discussed by McConnachie et al. (2009; 2010). If the stars in this 
structure are tidal in origin then they presumably 
were pulled from the disk during the most recent interaction with M31. In fact, 
simulations of interactions between M31 and M33 discussed 
by McConnachie et al. (2009) produce structures that are 
similar to those detected near M33, while also predicting significant warping of 
the M33 disk. However, McConnachie et al. (2010) measure a 
mean metallicity [Fe/H] $\sim -1.6$ for RGB stars in this feature, and 
such a low metallicity argues ostensibly against a disk origin. Still, if a 
large population of RGB stars with ages less than a few Gyr are present -- as 
would be expected if these stars originated in a star-forming disk environment -- 
then the metallicity estimated from color information will be skewed to low values 
if the calibration assumes that the RGB stars are all `old'.

	The distribution of sources with magnitudes and colors that fall within 
the 100 Myr boundaries defined in Figure 1 of Davidge et al. (2011) is 
shown in Figure 17. The source distribution is more-or-less uniform 
over large fractions of the northern and southern fields. A visual 
examination of individual sources in the 100 Myr sample indicate that many 
of the objects in these fields are either background galaxies 
or individual star-forming regions in the spiral arms of 
moderately distant spiral galaxies that evaded the culling described in \S 3.

	The substantial contamination by background galaxies notwithstanding, 
stellar groupings that are associated with the northern and southern 
end of the M33 disk are evident. Of particular note is 
the distinct spur of objects that extends to the north of the spiral arm, which 
will be referred to as the Northern Spur. Similar features do not eminate 
from the southern portion of the M33 disk.

	The CMDs of sources in the Northern Spur and the Northern Disk, 
where the latter region is identified in Figure 17, are 
compared in Figure 18. These structures have very different stellar contents, 
in that the Northern Spur lacks the very bright main sequence stars that populate 
the Northern Disk. The majority of blue sources in the Northern Spur have 
$u' > 23.5$, which corresponds to ages in excess of 100 Myr. The Northern Spur 
thus does not host obvious on-going star formation. Still, the Northern Spur may 
not be devoid of star-forming material, as the Putman et al. (2009) HI map shows 
a nose in the HI emission map that coincides with this structure.

	The outer regions of M33 thus may contain at least one stellar 
grouping -- the Northern Spur -- that coincides with a region of HI emission. 
Given the evidence for an interaction with M31 as recently as 1 Gyr in the past, 
then the Northern Spur might be the remnant of a structure that formed 
due to the interaction, and in \S 6.5 it is demonstrated that 
it is part of a much more diffusely distributed structure. Structures of this 
nature may dissipate over a few disk rotation timescales, or $\sim$ Gyr.

\subsection{Mapping Very Diffuse Structures}

	There are indications that extremely diffuse intermediate 
age structures lurk in the peripheral regions of M33. 
While contamination from background galaxies complicates 
efforts to identify such objects, diffuse stellar aggregates can still be detected 
using source counts that are averaged over large angular scales. Such structures 
are hard to detect in unsmoothed star counts. 

	The distribution of 100 Myr objects in all five fields is shown 
in Figure 19. These images show star counts in $500 \times 500$ MegaCam 
pixel gathers. This bin size corresponds approximately 
to $0.5 \times 0.5$ kpc, and so is conducive to the detection of objects 
with spatial scales of a kpc or higher. A low pass filter was applied 
to the raw number counts to suppress isolated concentrations of objects that occur 
on scales of a few hundred MegaCam pixels, as visual examination indicates that 
many of these tend to be star-forming knots in moderately distant background 
galaxies.

	In addition to the main body of the M33 disk, four distinct structures 
are seen in Figure 19. One of these is a low density extension of the Northern 
Spur. The connection between this structure and the disk is examined further in 
Figure 20, which shows the distribution of sources in the NE and NW fields, 
again with $500 \times 500$ MegaCam binning. The plume that contains the 
Northern Spur can be traced over $\sim 25$ arcmin, or $\sim 7$ kpc.

	Three conspicuous extended structures 
are labelled DO (`Diffuse Object') 1 -- 3 in Figure 19. 
The surface density of DO 2 is high enough that it can be seen 
in the upper right hand corner of the 100 Myr distributions in Figure 14. 
The compact nature of DO 3 is suggestive of a background cluster of 
galaxies. However, such a rich galaxy cluster is not obvious in the MegaCam data, 
and it is significant that -- like DO 2  -- DO 3 is located close to structure 
in the HI isophotes shown in Figure 1 of Putman et al. (2009), which is 
reproduced in Figure 19.

	DO 1 is located near the middle of the NE$+$NW fields, and is due north of 
M33. It is evident from Figure 20 that DO 1 may be connected to the main body of 
the M33 disk and the Northern Spur by a diffuse stellar bridge, as the sources 
counts to the south and south east of DO 1 are higher than to the north or west. 
A tongue of objects also extends south of DO 1, pointing directly to 
the M33 disk. DO 1 is $\sim 0.5$ degree East of the ridgeline 
of the HI arm mapped by Putman et al. (2009), and is at the eastern edge of 
the area labelled `S2' in Figure 8 of McConnachie et al. (2010). That stars with 
ages of a few hundred Myr are seen near the McConnachie et al. (2010) structure 
raises the possibility that it may contain stars that span a range of ages. 

	The presence of blue stars with ages $\sim 100$ Myr suggests that 
star formation occured in DO 1 (as well as DO 2 and 3) 
in the not too distant past. DO 1 is spread over $\sim 400$ arcmin$^2$ 
($\sim 30$ kpc$^2$), and thus is comparable in extent to areas of 
UV emission in the outer regions of M83 (Thilker et al. 2005b). 
XUV disks occur in only a fraction of nearby galaxies (e.g. 
Zaritsky \& Christlein 2007), and the so-called Type I XUV disks, in which 
the UV emission is clumpy rather than smoothly distributed, are likely 
the result of spiral structure propogating through the outer regions of 
disks (Bush et al. 2010).

	While M33 does not have an XUV disk at the 
present day (Thilker et al. 2005a), there are hints that  
M33 may have experienced vigorous star-forming activity at large radii within the 
past few hundred Myr. There is a relative excess of stars that formed $\geq 
100$ Myr in the past at distances in excess of 8 kpc from the galaxy center (\S 
5), and azimuthal mixing of tidal streams may contribute to producing 
such a diffuse stellar component (e.g. Oh et al. 2008). 
This being said, at least some of the older stars in the 
peripheral parts of the M33 disk may not have formed {\it in situ}, but may have 
migrated there as a result of dynamical interactions. Kinematic heating by 
sources that are external to the disk but still part of the extended dark matter 
field of the galaxy might also affect the distribution of stars 
at large radii (e.g. de Jong et al. 2007). 

	If star formation in DO 1 occured {\it in situ} then M33 
may have had an XUV disk $\sim 100$ Myr in the past. 
Alternatively, DO 1 may be the remnants of a 
stellar tidal stream, and this is supported by the close proximity to the 
structure identified by McConnachie et al. (2010).
Deep imaging studies provide a means of distinguishing between these 
possibilities. Both the XUV and tidal source models predict that DO 1 should 
contain main sequence stars that are fainter than those found here. 
However, the XUV and tidal stream models predict different main sequence LFs. 
Whereas XUV activity should produce main sequence stars with a LF 
that follows that of a SSP, if DO 1 contains disk stars that span a range of ages 
then the main sequence LF should depart significantly from that of a SSP.

\section{DISCUSSION \& SUMMARY}

	Local Group galaxies are fundamental calibrators 
for stellar content investigations of more distant systems, many of which are 
late-type spiral galaxies. As the nearest Sc galaxy, M33 is thus an obvious 
target for understanding the pedigrees of more distant galaxies.
In this study, moderately deep wide-field images recorded with the 
CFHT MegaCam have been used to probe the recent 
star-forming history of the M33 disk and its immediate surroundings. 
Main sequence stars that formed within the past few hundred Myr are resolved 
throughout a significant fraction of the $\sim 4.5$ degrees$^2$ that we examine.

	Our study yields three main results. First, the LF of main 
sequence stars indicates that -- when averaged over large parts of the disk -- 
the SFR within the central 8 kpc of M33 has not changed with time during the past 
few hundred Myr. This in turn suggests that the disk has not been disturbed 
for at least this period of time, given that tidal interactions will affect 
the SFR by -- for example -- re-distributing star-forming material.

	Second, the LF of main sequence stars at distances in excess of 8 kpc from 
the galaxy center is not consistent with a constant SFR during recent epochs. 
In fact, the peripheral regions of M33 are not locations of star formation 
at the present day, and the projected density of very young 
main sequence stars shows a steep drop near R$_{GC} = 8$ 
kpc. This radial truncation in the young star-forming disk is a robust 
result, that is also seen in the distribution of UV light (Thilker et al. 2005). 

	Third, we have found that the outer regions of M33 harbor 
diffuse collections of blue objects with photometric properties that are 
consistent with those of main sequence stars with ages $\sim 100$ Myr. 
Some of these structures appear to emerge from the disk, while others 
may be isolated. These structures may be tidal in nature, and/or may be 
spatially extended areas where star formation occured {\it in situ}. The latter 
interpretation opens the possibility that while star formation 
at the present day is restricted to within 
$\sim 8$ kpc of the galaxy center, during intermediate epochs it may have 
occured over a much larger region. The appearance of M33 at this time would 
have been very different from today, especially in the UV.

\subsection{Evidence for an Interaction in the SFH, and the Long-Term Impact on M33}

	A centrally concentrated upswing in star-forming activity 
is one consequence of a tidal encounter with another galaxy. Tidal torques remove 
angular momentum from gas in the disk, causing it to migrate inwards. 
Elevated levels of star formation in the inner regions of the galaxy are 
then triggered as the gas density builds and cools. 
Pre-existing stars in the disk gain angular momentum via interactions with 
the infalling gas clouds, and move to larger radii, producing an excess of 
old and intermediate age stars in the outer disk (e.g. Younger et al. 2008). 
If there was a recent interaction with M31 then elevated levels of central star 
formation may have occured in M33 -- is there evidence of such activity in 
the resolved stellar content?

	If the disk of M33 is recovering from an uptick in star-forming activity 
that occured within the past few hundred Myr, which is the typical timescale 
for a starburst, then the SFR should decrease as one moves towards more recent 
epochs. Such a decrease in the SFR is expected 
as the gas supply is depleted by star formation 
and/or feedback. Surveys of M33 star clusters suggest that a large peak in 
star-forming activity may have occured $\sim 0.1$ Gyr in the past (Chandar 
et al. 1999). However, the excellent match between the observed LFs and 
constant SFR models within 8 kpc of the galaxy center
indicates that the global SFR has not changed during the past $\sim 0.2$ Gyr. 
In fact, given that interactions fuel elevated levels of star 
formation over roughly a disk rotation time ($\sim 0.3$ Gyr), then the 
results presented in this paper suggest that an interaction with M31 could not 
have occured during the past $\sim 0.2 + 0.3 = 0.5$ Gyr.

	Studies that examine older stellar generations provide additional 
constraints as to when an interaction between M31 and M33 may have occured. 
Williams et al. (2009) find that the SFR near the center of M33 
peaked during the past 0.5 -- 2.5 Gyr, at which time the SFR 
jumped by an order of magnitude above the level that prevailed 2.5 -- 10 Gyr in 
the past. The time at which this jump in the SFR occured falls within the 
constraints set by the present-day positions and motions of M31 and M33 (Putman 
et al. 2009). Williams et al. (2009) also find that (1) the SFR varied with 
radius, as expected if gas is displaced by an interaction, and (2) the 
relative amplitude of the increase in the SFR during this same time interval drops 
with increasing R$_{GC}$, also as expected in the context of inward radial gas 
flows. 

	The global SFR during the past few 
hundred Myr has been lower than the long-term average (Williams et al. 2009), and 
the sSFR of M33 is lower than in most other late-type galaxies (Munoz-Mateos et 
al. 2007). These results are consistent with the depletion of cool gas 
throughout M33, as might occur if there was sustained large-scale star formation 
during intermediate epochs. The substantial jump in the central SFR during 
intermediate epochs and the subsequent drop in 
the SFR since that time suggests that M33 may have 
experienced a modest starburst; however, any starburst activity 
was not so dramatic as to present-day restrict star-forming activity to the 
central few kpc, as is seen in M82, where a large fraction of the outer 
stellar disk has been denuded of gas.

	The drop in the infall rate 3 Gyr in the past infered from the 
chemical enrichment properties of the outer disk of M33 
by Barker \& Sarajedini (2008) may be understood in the context of the 
stripping of the circumgalactic M33 HI reservoir by M31. The drop in infall 
occured at a time that is consistent with the interaction timescales estimated 
by Putman et al. (2009). The disruption of such a circumgalactic reservoir will 
affect the overall appearance of M33 in two ways. 

	First, the removal of gas from the outer halo of a 
galaxy may affect the size of the star-forming disk. Models of ram pressure 
stripping indicate that star-forming disks contract as the circumgalactic 
gas reservoir that fuels star-forming material is removed (e.g. 
Bekki 2009). Of course, the Local Group environment is such that M33 has not been 
subjected to significant ram pressure stripping. Still, if the 
M33 gas reservoir was disrupted tidally then 
the net impact on the star-forming disk may be similar to that produced by 
ram pressure stripping: the appearance of a compact star-forming disk for an 
extended period of time (presumably a significant fraction of 
the Hubble time) until an external gas reservoir is restored. The 
sharp outer disk boundary at 8 kpc in M33 may be a consequence of a truncated 
angular momentum content in the circumgalactic gas reservoir. 

	Second, the removal of an external reservoir that replenishes disk 
gas will change the SFR, and hence the appearance of the galaxy. 
If the SFR during the past few hundred Myr has been 0.7 
M$_{\odot}$ year$^{-1}$ (Blitz \& Rosolowsky 2006), then 
$0.7 \times 0.2 \times 10^9 = 1.4 \times 10^8$ M$_{\odot}$ of stars have 
formed in the past 0.2 Gyr. Adopting a 5\% star formation efficiency 
(Inoue et al. 2004), then this level of star formation requires $2.8 \times 
10^9$ M$_{\odot}$ of gas. This is sustantially lower than the present-day HI mass 
of M33 (Putman et al. 2009). To be sure, gas is recycled throughout the disk; 
still, the point to be made is that M33 is not gas-rich and the SFR, which 
has been constant for the past few hundred Myr, will gradually drop as the 
disk gas supply is depleted. The drop in SFR during the past 10 Myr that is 
predicted by the star counts in \S 4.2 may signal the start of this process.

\subsection{The Distribution of Stars and Gas in M33 as a Probe of Interaction Timing}

	The radial distributions of gas and stars provide clues into the evolution 
of disks. Gas and stars are expected to have different distributions in isolated 
disks. This is due in part to the disk assembly process, as the angular momentum 
distribution of accreted gas will differ from that of 
stars that are already present. However, even in the absence of 
gas accretion, the distinct dynamical behaviours of stars and 
gas results in different radial distributions. Because of their small 
cross-sections when compared with typical stellar separations, stars form 
collisionless systems, while the relatively large 
cross-sections of gas clouds make them collisional systems. 
Stars may also evolve into extraplanar systems that are distinct from the 
gas distribution, even in isolated galaxies. Indeed, 
some fraction of disk stars at small radii may evolve into 
a thick disk, as a natural consequence of inside-out disk formation
(e.g. Schonrich \& Binney 2009). 

	The distributions of gas and stars in disks have been modelled, and 
the results can be compared with the observed distributions in M33. Here, we 
consider the cosmologically-based galaxy simulations examined by Martinez-Serrano 
et al. (2009). A break in the model light profile 
typically occurs at $\sim 3$ disk scale lengths. 
The gas distribution in these models is roughly flat within the 
break radius, and gradually decreases beyond this point. Star formation 
continues past the break radius, occuring out to 4.3 scale lengths, at 
which point it ceases due to the low density of star-forming material. 
The ratio of gas mass to stellar mass peaks at the break radius, making this
the point in the disk with the lowest photometrically-weighted age. 

	The HI density profile throughout much of the M33 disk is relatively flat 
(Corbelli 2003; Putman et al. 2009), in agreement with the 
Martinez-Serrano et al. (2009) models. The general trend of increasing specific 
SFR with radius in the main body of the disk predicted by the models is also 
consistent with the relative contributions made by young and old stars shown in 
Figure 15. Still, the muted evidence for star formation outside 
of the break radius in M33 is not consistent with the models, 
and the presence of tidal features in the HI distribution outside 
of the disk (Putman et al. 2009) points to an event that undoubtedly affected 
the evolution of M33. The presence of large-scale tidal 
features aside, the distributions of gas and stars 
within the break radius of M33, where the characteristic timescale for 
relaxation is shorter than at large radii, are at least qualitatively consistent 
with predictions made for disk galaxies evolving in the 
CDM cosmological paradigm. That the relative distributions 
of gas and stars in the main disk of M33 are consistent with that of an 
isolated disk suggests that sufficient time has passed 
(i.e. a few disk rotations) to allow the distributions of stars and gas 
within the disk to regain at least the semblance of an equilibrium state.

\subsection{The M31-M33 Connection}

	An interaction between M31 and M33 would almost certainly have imprinted 
signatures in the stellar content of the former. McConnachie et al. (2009) 
suggest that an interaction with M33 would have disturbed the outer disk of M31, 
causing a warp and displacing disk stars to large radii. An interaction 
with M33 may also have triggered an inward flow of gas in M31, depleting 
gas reserves at large radii. Indeed, the gas content of M31 at large radii is 
deficient when compared with models of isolated evolution (Yin 
et al. 2009). The interaction may also have produced elevated SFRs in 
M31, and a sharp drop in overall star-forming activity would occur as gas is 
consumed during star formation or is disrupted by feedback. Williams 
(2002) concludes that a global downturn in star-forming activity occured in M31 
$\sim 1$ Gyr in the past. Given that starbursts appear to last no more than a Gyr 
(e.g. Leitherer 2001), then this result -- if due to an interaction with M33 -- 
suggests that M31 and M33 interacted no more than $\sim 2$ Gyr ago, which is 
consistent with the age ranges estimated by Putman et al. (2009) 
and McConnachie et al. (2009). Of course, all of these 
properties of M31 could also be explained by an interaction with another 
(possibly now defunct) companion.

	If there was a close interaction between M31 and M33 then some of the 
features traditionally associated with M31 companions could instead be remnants of 
this interaction. Consider the M31 Giant Southern 
Stream (GSS; Ibata et al. 2001). The simulations run by Bekki (2008) 
indicate that M33 may leave a gaseous debris stream near M31, and Fardal et al. 
(2008) conclude that the GSS may have originated from a spiral galaxy. Indeed, 
the GSS contains stars with a metallicity as high as [Fe/H] $\sim -0.5$ 
(Guhathakurta et al. 2006), which is consistent with an origin in the M33 disk. 
The velocity dispersion of stars in the GSS (e.g. Ibata et al. 
2004) is also consistent with an origin in a kinematically 
cold environment, such as a disk. Finally, the GSS is 85 kpc behind M31 
(e.g. Tanaka et al. 2010), and simulations suggest that the 
GSS was produced by an object on a highly eccentric orbit (Font et 
al. 2006; Fardal et al. 2006; 2007), such as M33.

	While some properties of the GSS can be rationalized in the context 
of an interaction with M33, efforts to link the GSS with such an event 
run into timing issues. The GSS has a dynamical age of a few hundred 
Myr (e.g. Font et al. 2006), which is more recent than the minimum probable age 
for an interaction estimated by Putman et al. (2009), and is not consistent with 
the time interval during which there has been a constant SFR throughout 
the M33 disk. Another issue is that the vast majority of 
stars in the GSS have an age $\geq 4$ Gyr (Brown et al. 
2006), and so fall outside of the nominal age range predicted 
for an interaction. Still, a modest population of 
young stars is present in the GSS (e.g. Figure 17 of Brown et al. 2006). 
Recently formed stars in extraplanar environments may not be well-mixed with 
underlying stellar material, as they formed in gas concentrations, whereas 
pre-existing stars will tend to dissipate spatially. If this is the case in M31 
then future observations may find concentrations of intermediate age stars in 
different fields than those sampled by Brown et al. (2006).

\parindent=0.0cm

\noindent

\clearpage

\begin{table*}
\begin{center}
\begin{tabular}{lll}
\tableline\tableline
Program ID & $u*$ & $g'$ \\
\tableline
2003BF15 & 715153--156 & 714745 \\
 & & 715150, 797, 935, 938 \\
 & & 716134 \\
 & & 718659, 894 \\
 & & 719309, 585 \\
 & & 724409 \\
 & & 727548 \\
2004BH06 & 774552 & \\
 & 774554--557 & \\
2004BH20 & & 757867, 868 \\
 & & 762372, 373, 374, 378, 380, 381, 382, 466--472 \\
2004BF26 & & 759110 \\
 & & 761403, 560 \\
 & & 765027, 128, 927 \\
 & & 767646, 804, 809, 810, 975 \\
 & & 769086, 541 \\
 & & 773381 \\
\tableline
\end{tabular}
\end{center}
\caption{Archival Data Used to Construct the Center Field Images}
\end{table*}

\clearpage

\begin{figure}
\figurenum{1}
\epsscale{0.75}
\plotone{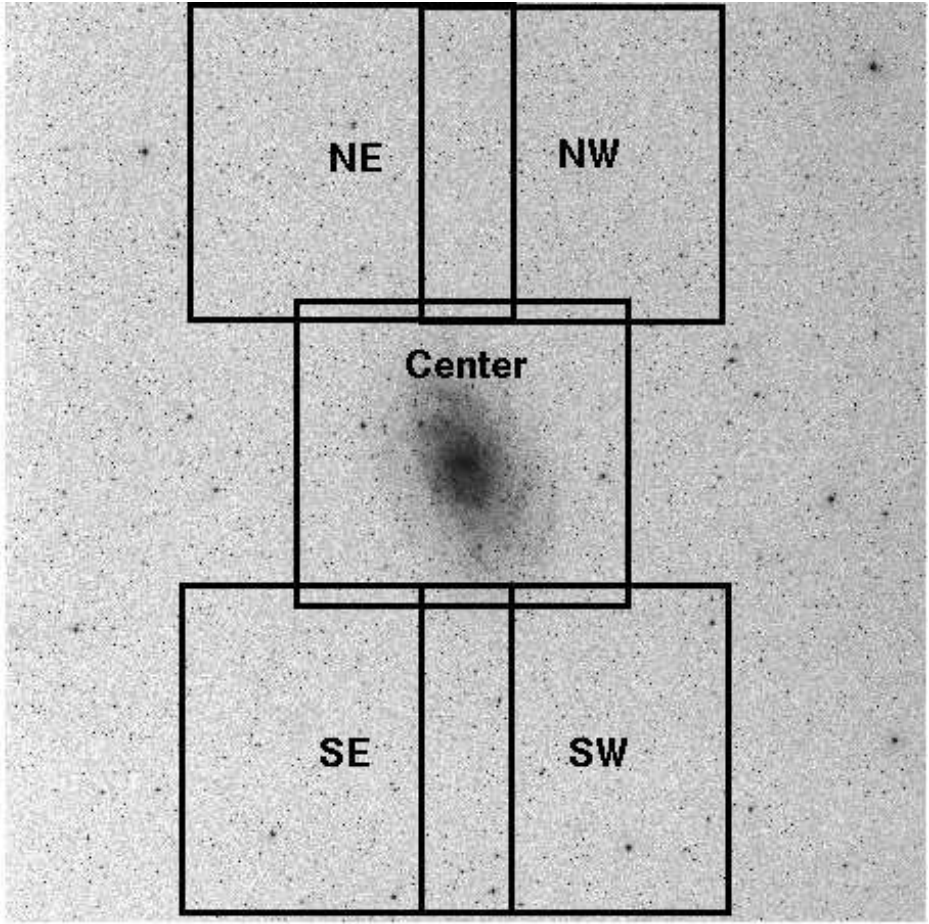}
\caption{The locations of the fields discussed in this paper. The underlying 
reference image is from the DSS in the blue filter. East is to the left 
and north at the top. Each MegaCam field covers roughly 1 degree$^2$, 
and is labelled with the identifier that is used throughout the paper.}
\end{figure}

\clearpage

\begin{figure}
\figurenum{2}
\epsscale{0.75}
\plotone{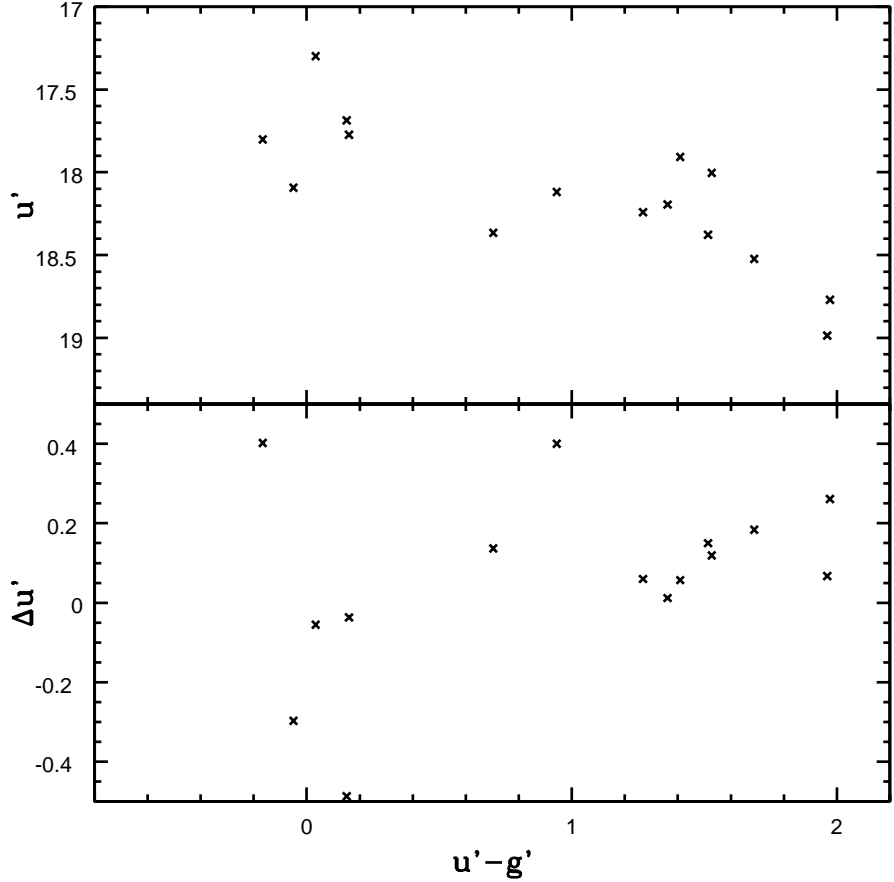}
\caption{(Top panel:) The CMD of bright, isolated sources that are 
also in the sample observed by Massey et al. (2006). The majority of 
objects in both samples have red colors, and this is because the crowded regions 
of the disk, where bright blue stars are located, were intentionally 
avoided when comparing the two sets of measurements. 
(Lower panel:) The difference between the MegaCam and NOAO 
measurements, $\Delta u'$, with respect to $u'-g'$. The Spearman correlation 
coefficient indicates that $\Delta u'$ does not vary in a systematic way with 
color. With the exception of the four outliers, 
the majority of points have $\pm 0.1$ magnitude scatter in $\Delta u'$, 
which is comparable to the dispersion about the empirical relation between 
$u'-g'$ and $U-B$ in Figure 13 of Smith et al. (2002).}
\end{figure}

\clearpage

\begin{figure}
\figurenum{3}
\epsscale{0.75}
\plotone{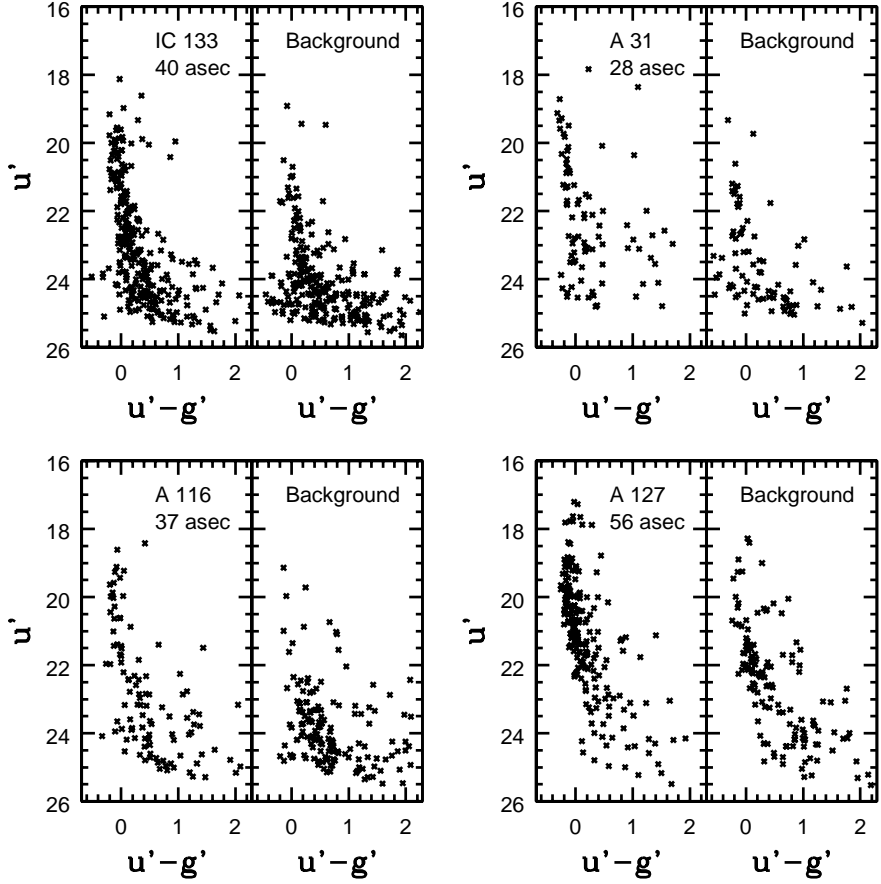}
\caption{The $(u', u'-g')$ CMDs of four star-forming complexes that are 
located in low-density regions of the M33 disk. The approximate radial size of 
each cluster is specified. The CMDs of sources in an annulus that surrounds each 
cluster are also shown; this annulus samples the same total area on the sky 
as the cluster that it abuts. The main sequence is clearly seen in 
each CMD, and sources to the right of the main sequence are BSGs. 
The CMDs of IC 133, A 31, and A 116 are well-populated down to $u' = 24 - 25$. 
However, the number counts in the A 127 CMD drop when $u' > 22$. 
This drop is probably due to crowding, as A 127 is the structure with the 
highest mean surface brightness. The objects with $u' > 22$ in the A 127 
CMD are located in lower density parts of this complex.}
\end{figure}

\clearpage

\begin{figure}
\figurenum{4}
\epsscale{0.75}
\plotone{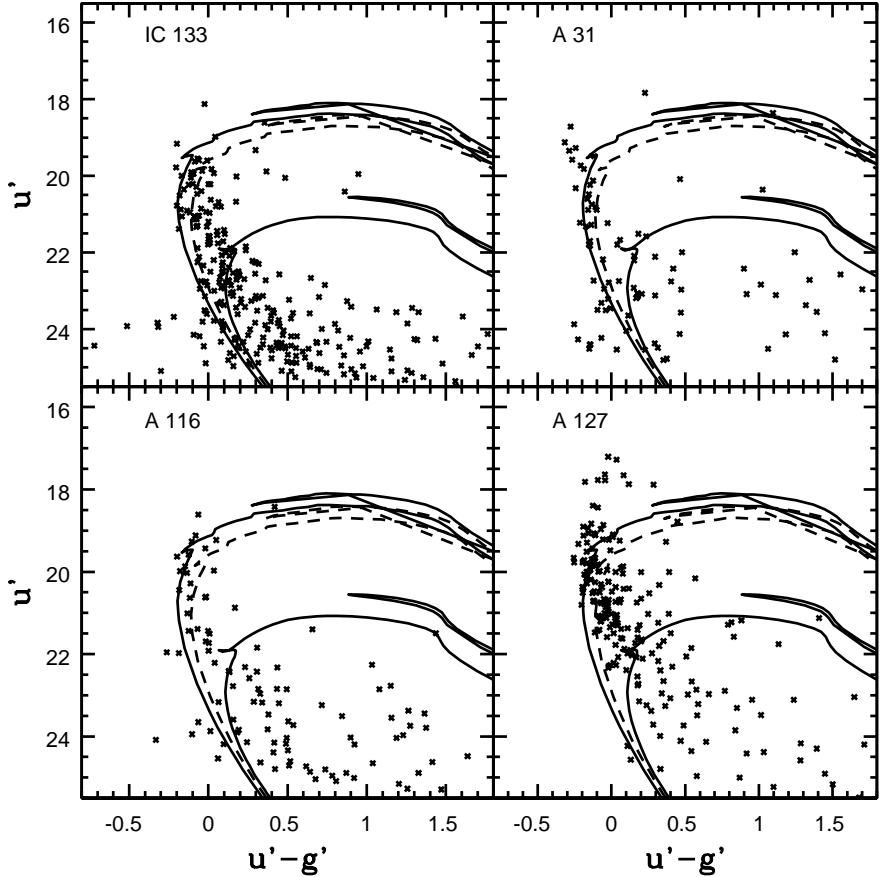}
\caption{The $(u', u'-g')$ CMDs of four star-forming complexes are compared 
with isochrones from Girardi et al. (2004). The models have Z = 0.008 and ages 
10 Myr and 40 Myr. The isochrones plotted as solid lines assume a 
combined foreground and internal extinction A$_B = 0.34$ 
magnitudes, which is the default reddening used throughout this study (\S 1). 
The dashed line shows the 10 Myr isochrone with an additional extinction A$_V = 
0.2$ mag. The CMD of A 31 is well matched by the 10 Myr isochrones down to $u' 
\sim 24$, whereas the CMD of A 116 follows the 10 Myr isochrone when $u' < 22$. 
However, the agreement with the isochrones is not as good for 
IC 133 and A 127. In the case of IC 133 the morphology of the CMD 
suggests that this structure might contain stars that formed within the past 
$\sim 100$ Myr, which is the approximate lifetime for large star-forming 
complexes (e.g. Davidge et al. 2011). The agreement with the models is poorest 
for A 127.}
\end{figure}

\clearpage

\begin{figure}
\figurenum{5}
\epsscale{0.75}
\plotone{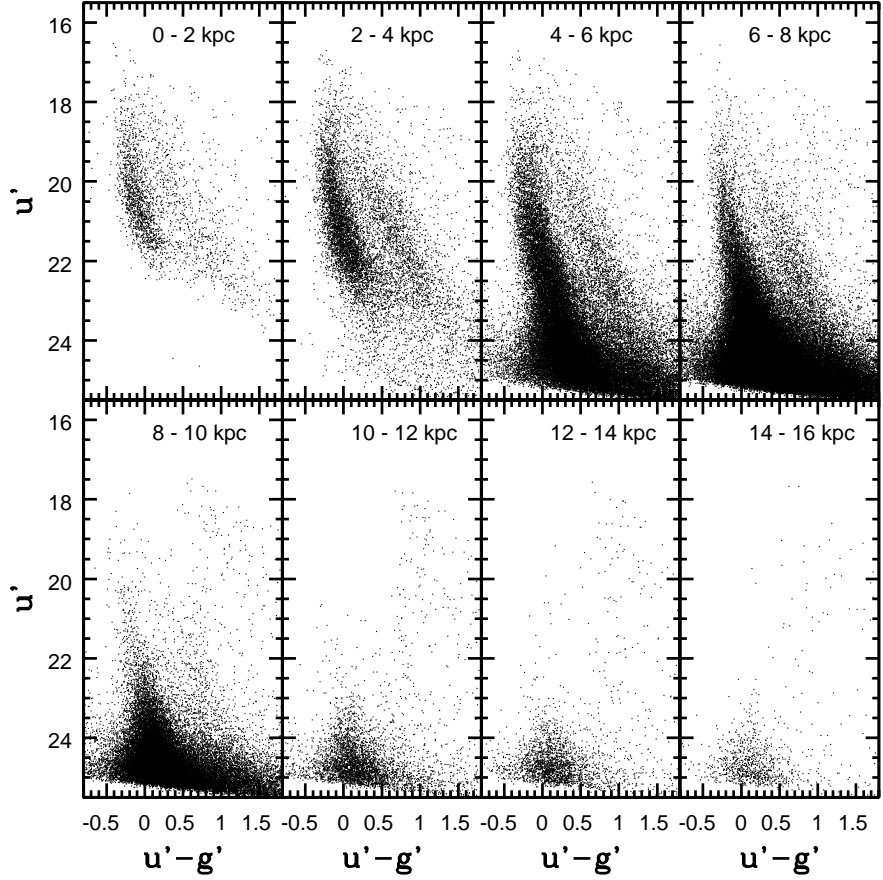}
\caption{The $(u', u'-g')$ CMDs of stars in the disk of M33. 
A bright main sequence dominates the CMDs with R$_{GC} \leq 10$ 
kpc, and the peak brightness of the bluest objects is roughly constant when 
R$_{GC} \leq 8$ kpc. BSGs form a distinct spray of stars to the right of the 
main sequence. The plume of objects with $u'-g'$ between 1.0 and 1.5 in the 
CMDs with R$_{GC} \geq 8$ kpc is populated by background galaxies.} 
\end{figure}

\clearpage

\begin{figure}
\figurenum{6}
\epsscale{0.75}
\plotone{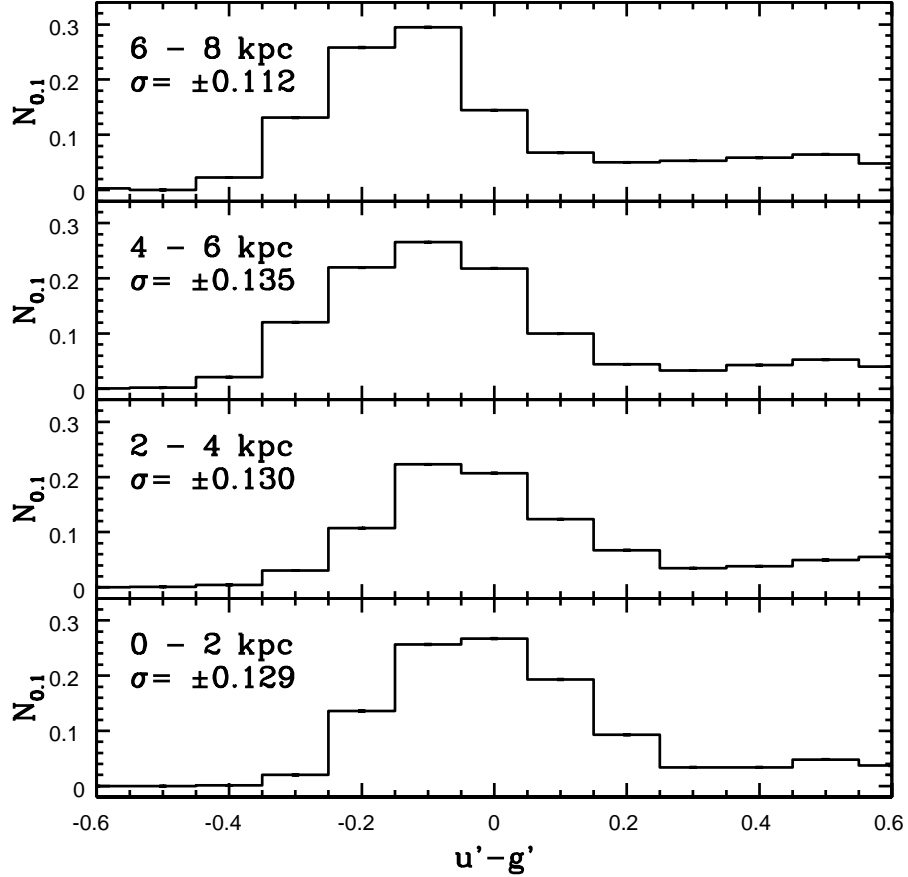}
\caption{The $u'-g'$ distributions of stars with $u'$ between 
20.5 and 21.5. The prominent peak in each color distribution is 
due to main sequence stars, while the tail to the right of this peak 
is made up of stars that have evolved off the main sequence. 
The distributions have been normalized according to the number of 
objects with $u-g'$ between --0.35 and 0.35; $\sigma$ is the standard deviation 
of each distribution. There is a trend towards bluer mean $u'-g'$ colors in 
the main sequence as R$_{GC}$ increases. If the SFH 
in the past few tens of Myr has not changed, as suggested by the investigation of 
main sequence LFs in \S 5, then this color trend may be indicative of higher 
levels of internal extinction at smaller R$_{GC}$, such that 
$\Delta A_V/\Delta R_{GC} = -0.05$ mag kpc$^{-1}$. The $u'-g'$ distributions of 
main sequence stars are wider than predicted from random photometric errors, and 
the excess width suggests that differential reddening within the inner disk of 
M33 introduces a dispersion $\Delta E(B-V) \leq \pm 0.10$ mag.}
\end{figure}

\clearpage

\begin{figure}
\figurenum{7}
\epsscale{0.75}
\plotone{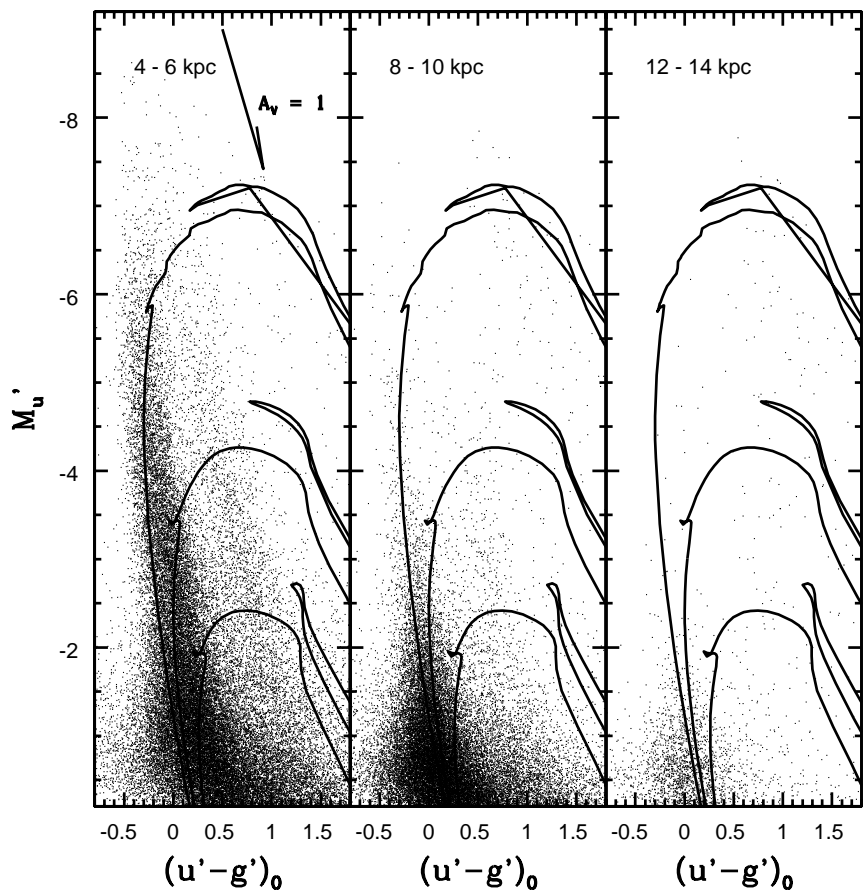}
\caption{The $(M_{u'}, u'-g')$ CMDs of sources in three annuli. 
Also shown are Z = 0.008 isochrones from Girardi et al. (2004) with 
ages 10 Myr, 40 Myr, and 100 Myr. There is reasonable agreement 
between the isochrones and the observed locus of main sequence stars, lending 
credence to the baseline reddening value. A reddening vector with 
a length that corresponds to A$_V = 1$ mag is shown in the 4 -- 6 kpc panel.}
\end{figure}

\clearpage

\begin{figure}
\figurenum{8}
\epsscale{0.75}
\plotone{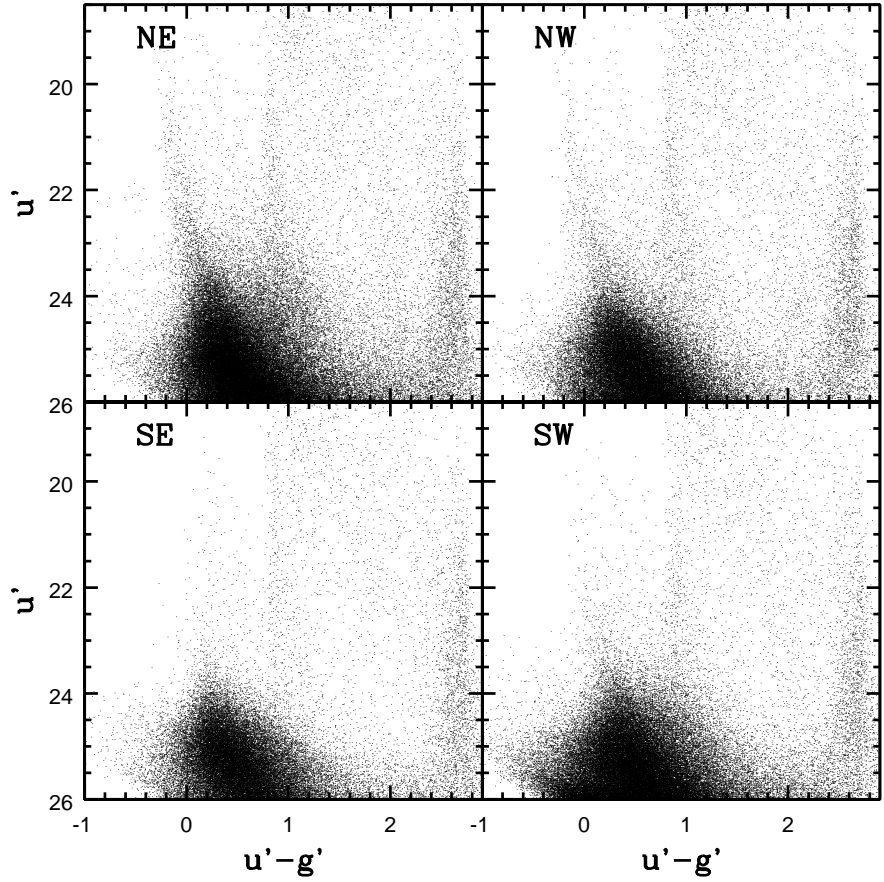}
\caption{The $(u', u'-g')$ CMD of sources in the NE, NW, SE, and SW fields. 
The bright blue plume in the NE and NW CMDs is populated by massive main 
sequence stars that are near the edge of the northern spiral arm. 
A less well-populated main sequence is seen 
in the CMD of the SW field. The majority of sources with $u'-g' > 1$ 
and $u' > 24$ are background galaxies, although the evolutionary tracks of 
intermediate age He-burning stars also pass through this part of the CMD.}
\end{figure}

\clearpage

\begin{figure}
\figurenum{9}
\epsscale{0.75}
\plotone{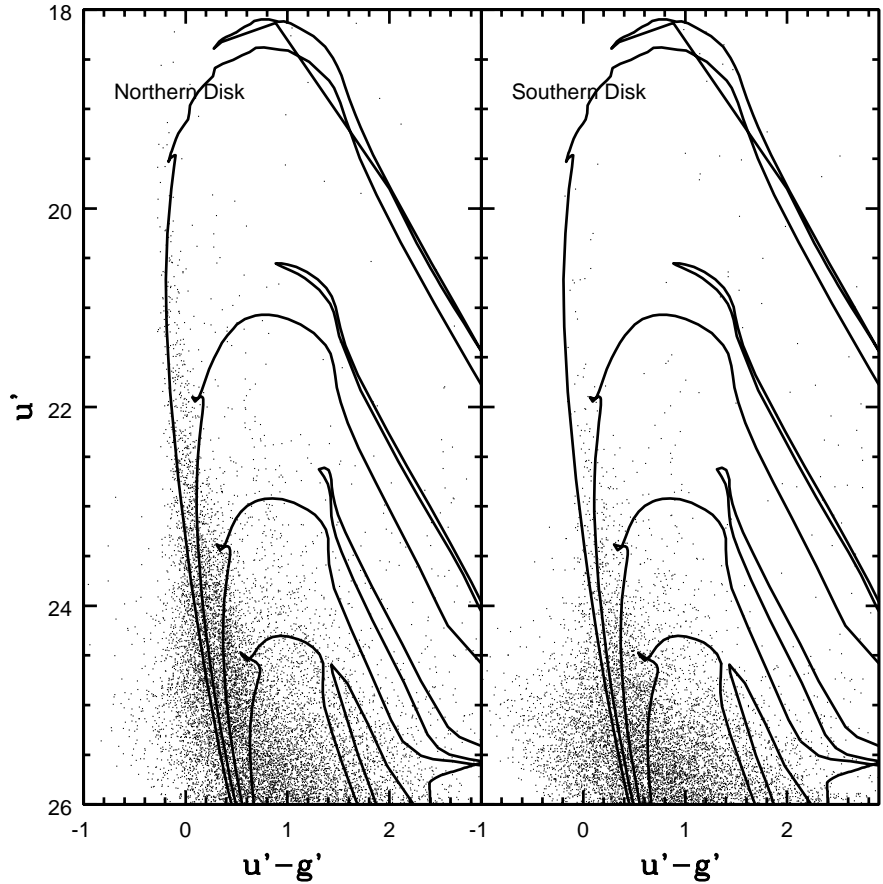}
\caption{The $(u', u'-g')$ CMD of sources at the periphery of the M33 
disk, in the areas indicated in Figure 17. The CMDs are 
compared with Z=0.008 isochrones from Girardi et al. (2004) that 
have ages 10 Myr, 40 Myr, 100 Myr, and 200 Myr. 
The blue envelope of upper main sequence stars in the northern 
disk follows the 10 Myr isochrone, and the observed main sequence width 
matches that predicted by the models. In contrast to the Northern Disk, only a 
handful of stars in the Southern Disk formed within the past few tens of 10 Myr.}
\end{figure}

\clearpage

\begin{figure}
\figurenum{10}
\epsscale{0.75}
\plotone{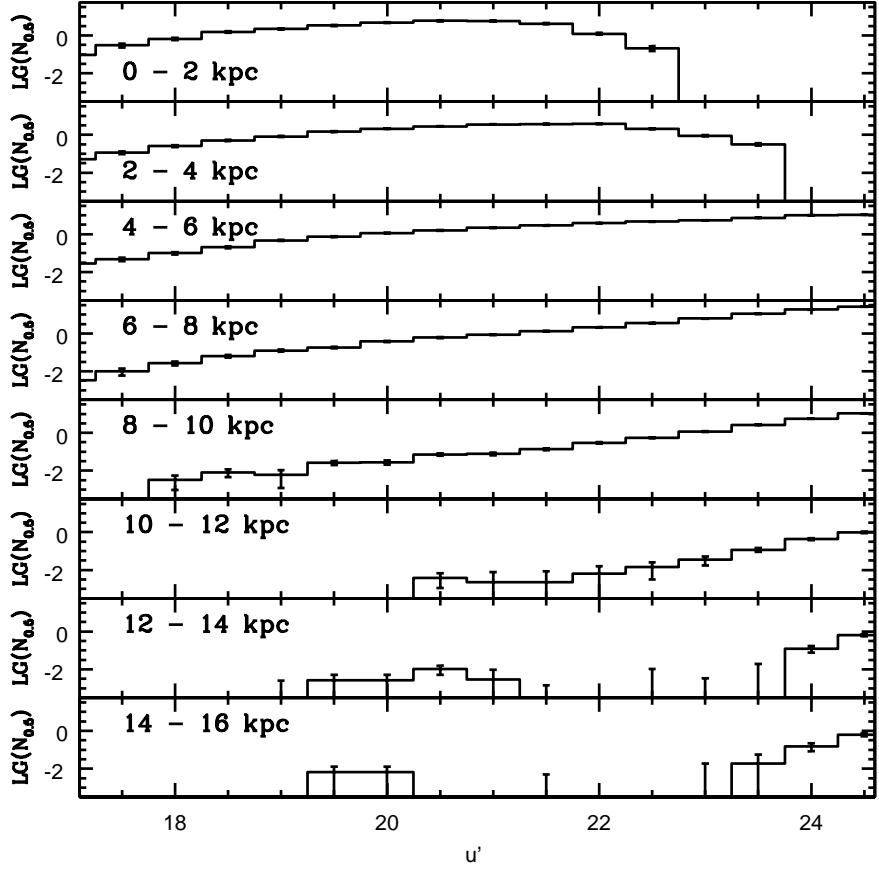}
\caption{The LFs of main sequence stars in the Center Field. Distances 
are measured in the plane of the M33 disk, and N$_{0.5}$ is the number of sources 
arcmin$^{-2}$ per 0.5 magnitude interval in $u'$, corrected for foreground 
star/background galaxy contamination using source 
counts from the outermost parts of the NE field. 
Crowding elevates the faint limit in the 0 - 2 and 2 - 4 kpc intervals 
when compared with larger R$_{GC}$. The shape of the LFs changes near 8 kpc. The 
LFs with R$_{GC} \leq 8$ kpc follow power laws that can be characterized by a 
single exponent. In contrast, at larger radii the bright end of the LF flattens 
while the faint end steepens. The peak brightness of the main sequence also drops 
with increasing R$_{GC}$ when R$_{GC} \geq 8$ kpc. These results are consistent 
with previous studies that have found that the distribution of young stars 
in M33 changes near R$_{GC} = 8$ kpc.}
\end{figure}

\clearpage

\begin{figure}
\figurenum{11}
\epsscale{0.75}
\plotone{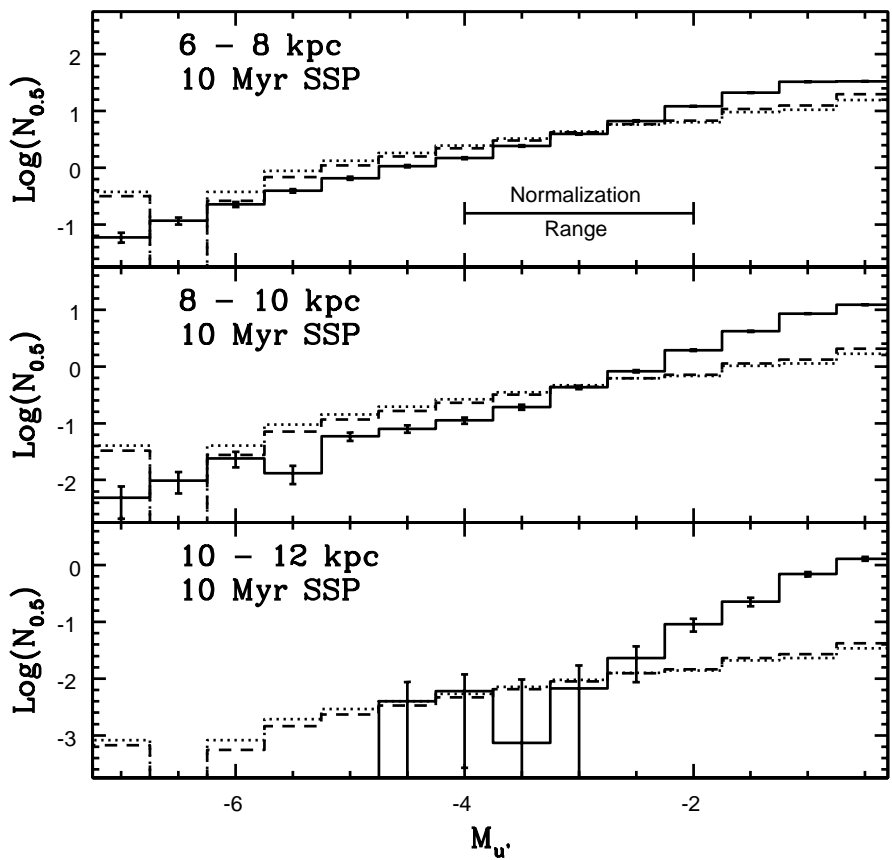}
\caption{Simple stellar population (SSP) model LFs with Salpeter (dotted line) and 
Kroupa (dashed line) IMFs are compared with observed LFs.
The models are normalized to match the observations in the interval 
M$_{u'} = -2$ to --4. N$_{0.5}$ is the number of sources arcmin$^{-2}$
per 0.5 magnitude interval in $u'$. The SSP model LFs are significantly flatter 
than the observed LFs, indicating that the youngest stars in the M33 disk did 
not form during a single event.}
\end{figure}

\clearpage

\begin{figure}
\figurenum{12}
\epsscale{0.75}
\plotone{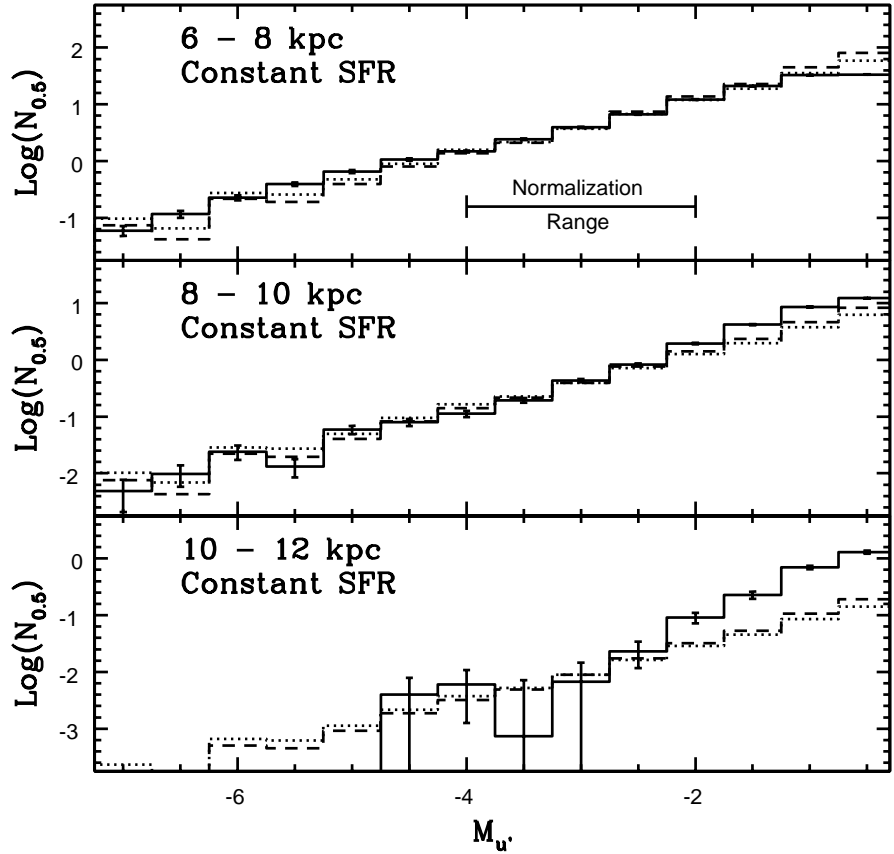}
\caption{The same as Figure 11, but showing constant SFR models. 
These models provide a reasonable match to the 6 -- 8 kpc and 
8 -- 10 kpc LFs over a wide range of magnitudes. However, the 10 -- 12 kpc LF is 
steeper than predicted by the models.}
\end{figure}

\clearpage

\begin{figure}
\figurenum{13}
\epsscale{0.75}
\plotone{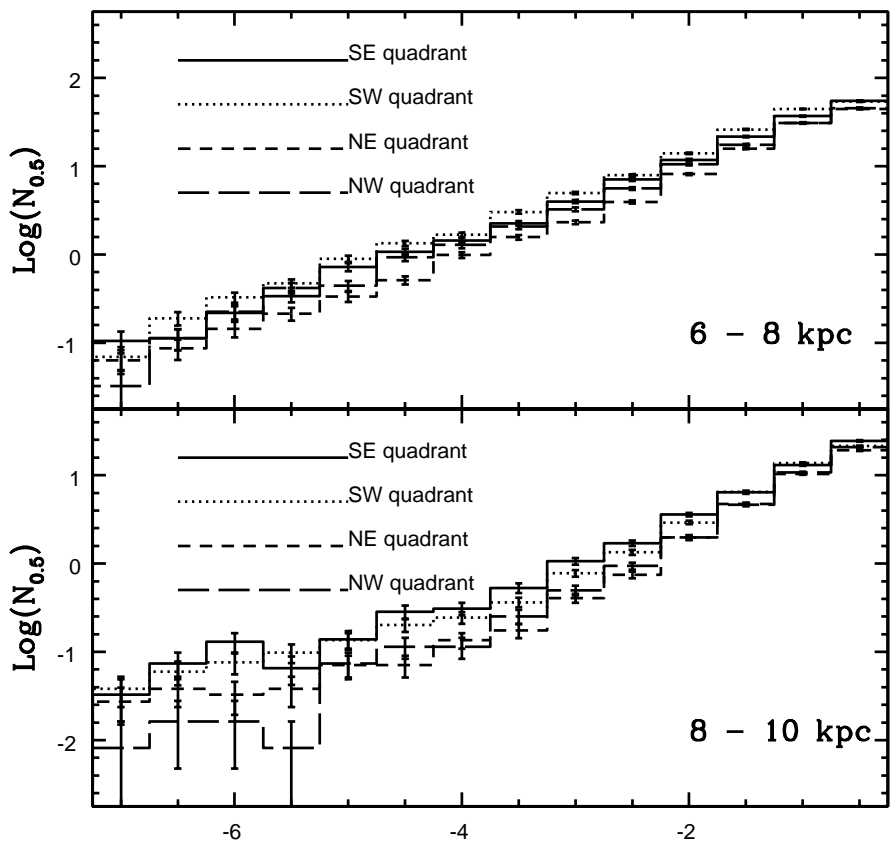}
\caption{Quadrant-to-quadrant variations in the LFs of main sequence stars are 
examined in this figure. N$_{0.5}$ is the number of sources arcmin$^{-2}$ per 0.5 
magnitude interval in $u'$ in each quadrant. There 
is a systematic tendency for the number counts in the interval M$_{u'} \leq -1$ 
in the southern half of the galaxy to be higher than in the northern half. 
This suggests that for R$_{GC} > 6$ kpc the southern half of the disk 
experienced a higher SFR during the past few tens of Myr than the northern half 
of the disk.}
\end{figure}

\clearpage

\begin{figure}
\figurenum{14}
\epsscale{0.75}
\plotone{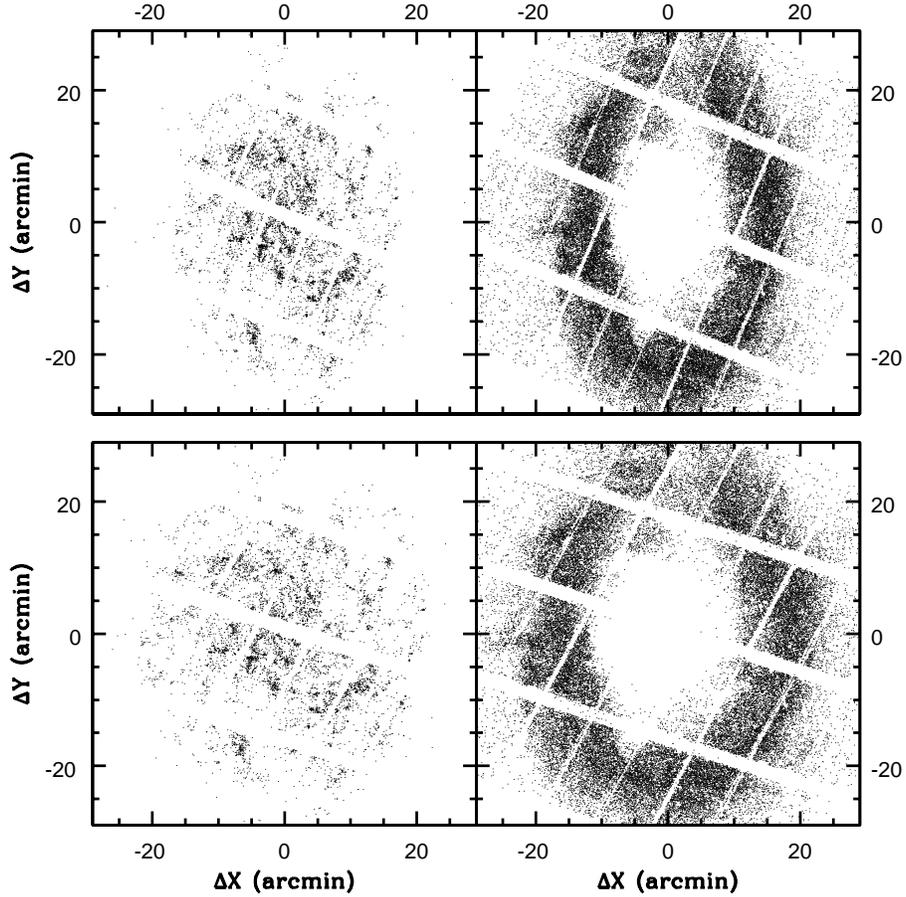}
\caption{The observed (top row) and de-projected (bottom row) distributions 
of stars in M33. The 10 Myr and 100 Myr samples were selected based on location 
in the $(u', u'-g')$ CMD, as defined in Figure 1 of Davidge et al. (2011). 
M33 is oriented such that the major axis parallels the vertical axis. 
The axis labels are offsets from the center of M33 in the rotated 
reference frame. The spiral structure defined by the 10 Myr data
can be traced into the central regions of the galaxy. 
With the exception of the sharp outer disk boundary, distinct structures are 
much less obvious in the 100 Myr sample. Crowding prevents the detection of main 
sequence stars with ages $\sim 100$ Myr in the central regions of the galaxy.} 
\end{figure}

\clearpage

\begin{figure}
\figurenum{15}
\epsscale{0.75}
\plotone{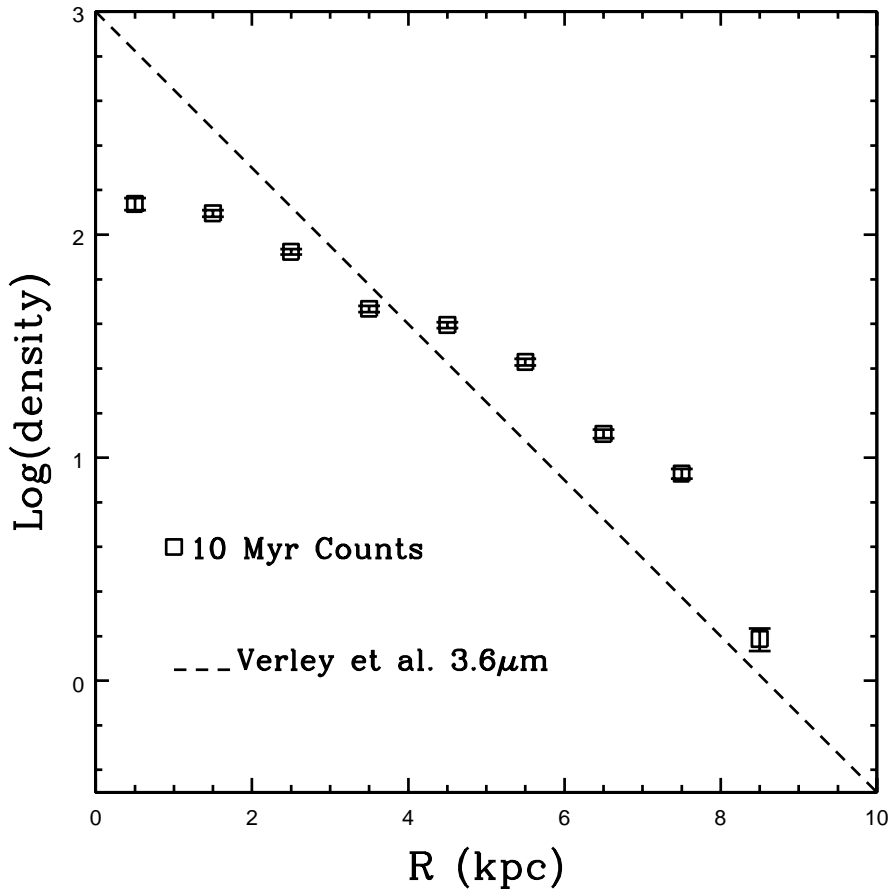}
\caption{The number of 10 Myr sources kpc$^{-2}$ is compared with the $3.6\mu$m 
integrated light profile from Figure 2 of Verley et al. (2009). The 
slope of the 3.6$\mu$m profile has been adjusted to account for the 
distance modulus adopted for the present study, which differs from 
that assumed by Verley et al. (2009). The light at $3.6\mu$m 
is dominated by old and intermediate-age stars, and so traces stellar 
mass. The 10 Myr number counts define a more-or-less linear 
relation out to R$_{GC} = 8$ kpc that is flatter than the 
$3.6\mu$m light profile. The contribution made by young stars to the total 
stellar mass density thus grows with increasing R$_{GC}$ out to 8 kpc, but 
then drops at larger radii.}
\end{figure}

\clearpage

\begin{figure}
\figurenum{16}
\epsscale{0.75}
\plotone{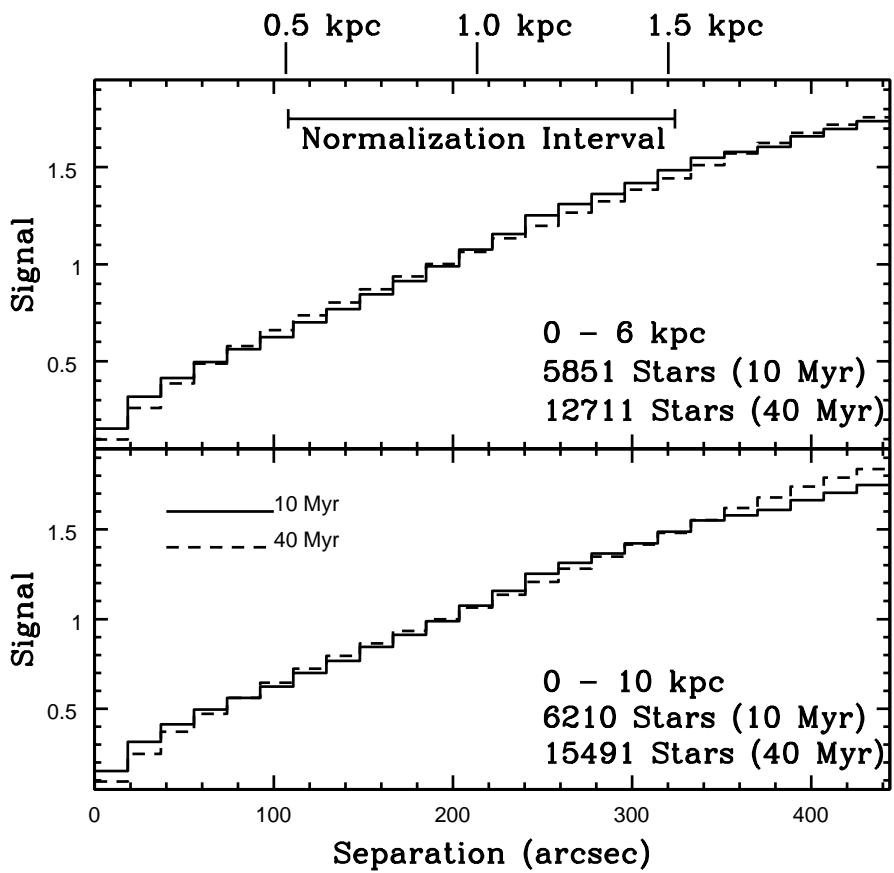}
\caption{The star-star separation functions (S3Fs) of main sequence stars in the 
10 and 40 Myr groups. The signals have been normalized to the numbers of sources 
with separations between 0.5 and 1.5 kpc (110 - 330 arcsec). 
At separations $< 200$ arcsec ($< 0.9$ kpc) the 40 Myr S3Fs 
are steeper than the 10 Myr functions. The signal at 
separations $< 60$ arcsec ($< 280$ parsecs) in both radial intervals of the 40 Myr 
S3Fs is then smaller than in the 10 Myr sample, indicating a diminished degree of 
grouping over these spatial scales.}
\end{figure}

\clearpage

\begin{figure}
\figurenum{17}
\epsscale{0.75}
\plotone{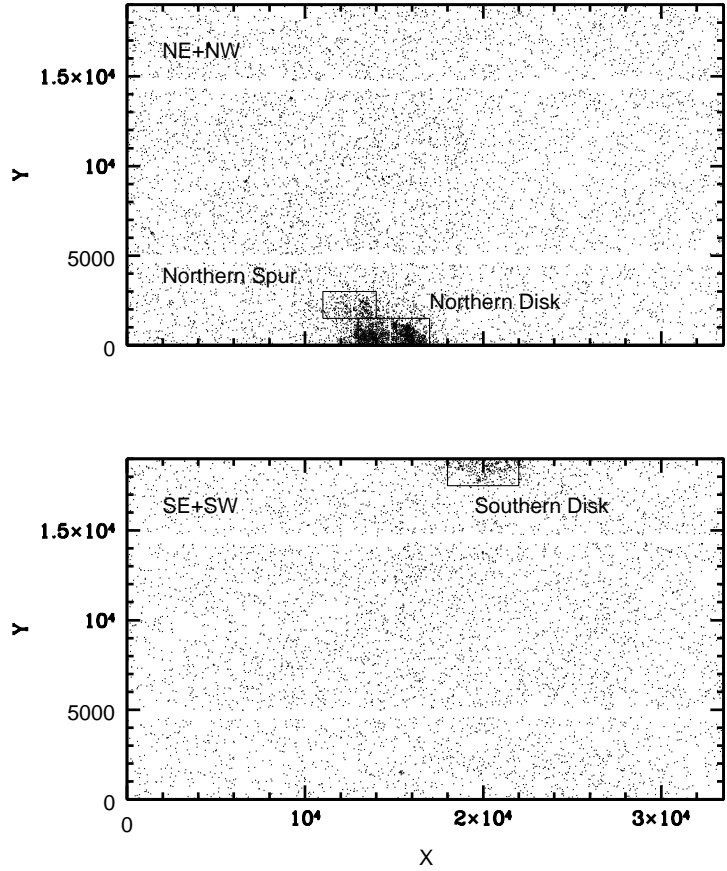}
\caption{The distribution of blue sources in the NE, NW, SE, and SW 
fields. The objects have magnitudes and colors that place them within the 100 Myr 
boundaries defined in Figure 1 of Davidge et al. (2011). While 
having photometric properties that are consistent with those of main sequence 
stars, many of these sources are in fact associated with background galaxies. 
However, stellar concentrations near the northern and southern ends 
of the disk are evident, and these are indicated.}
\end{figure}

\clearpage

\begin{figure}
\figurenum{18}
\epsscale{0.75}
\plotone{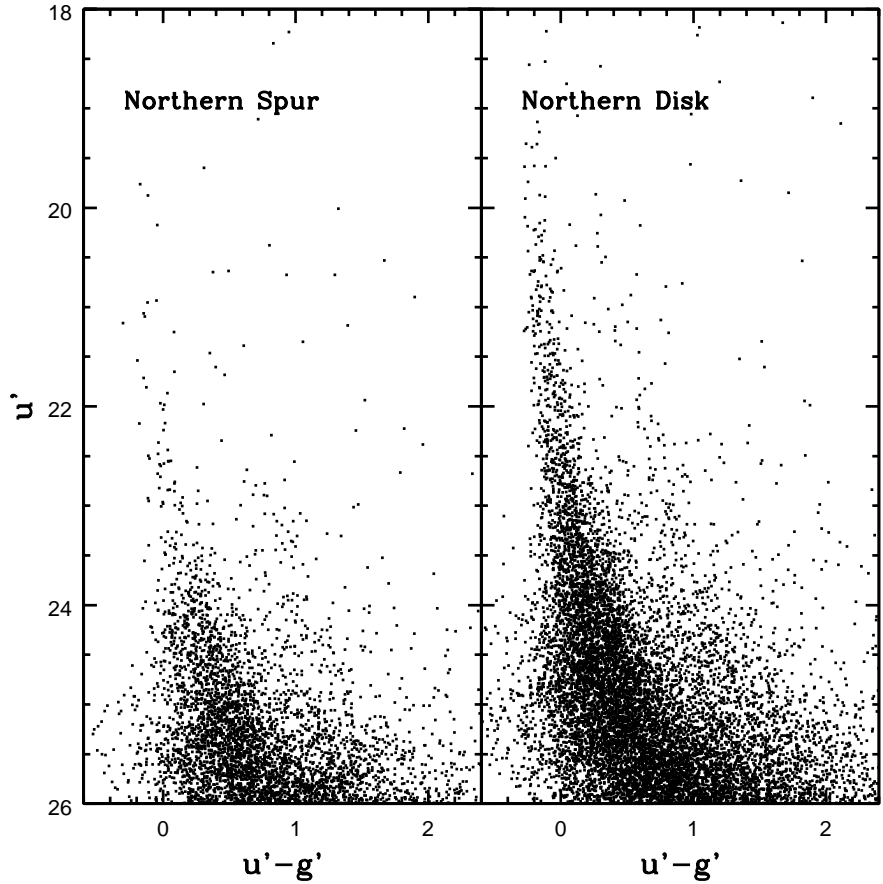}
\caption{The $(u', u'-g')$ CMDs of the Northern Spur and Northern Disk. The 
sources used to make these CMDs fall within the 
areas indicated in Figure 17. The Northern Spur lacks 
the very bright main sequence stars that are seen in the 
Northern Disk. The youngest stars in the Northern Spur tend to have ages of a few 
tens of Myr, although the Putman et al. (2009) HI map shows that this is an 
area of localized HI emission.}
\end{figure}

\clearpage

\begin{figure}
\figurenum{19}
\epsscale{0.75}
\plotone{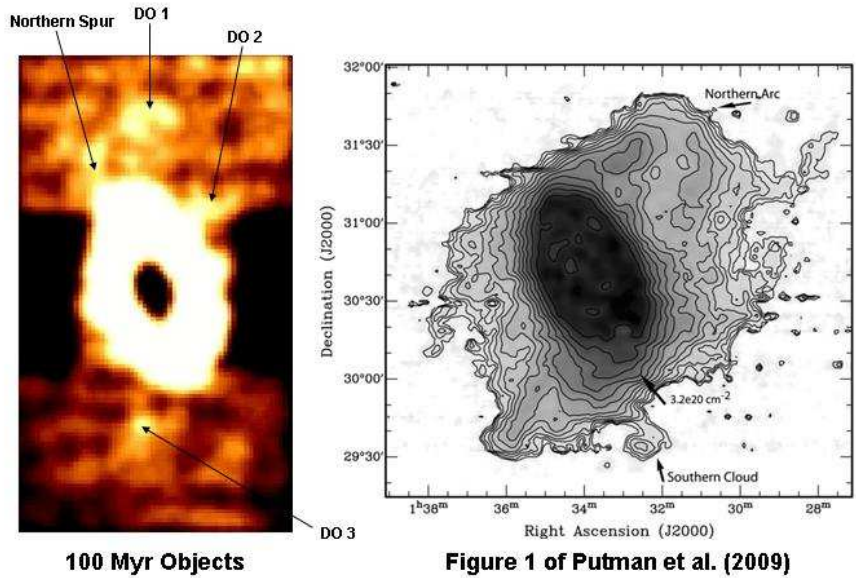}
\caption{(Left hand panel) The distribution of blue sources in the MegaCam fields. 
An area covering $3 \times 1.7$ degree$^2$ is shown, with north at the top, 
and east to the left. The intensity of each pixel reflects the number of sources 
in the 100 Myr sample in $500 \times 500$ Megacam pixel ($\sim 0.5 \times 0.5$ 
kpc) gathers, after applying a low-pass filter to remove star-forming regions in 
background galaxies. Structures that are discussed 
in the text are labelled. (Right hand panel) Figure 1 of Putman et al. (2009), 
which shows the HI distribution in and around M33. 
This figure is displayed with the same spatial scale as the MegaCam image 
to facilitate the cross-identification of features in the HI and MegaCam data. 
Note that the HI contours are not smooth in the vicinity of the structures 
that are identified in the left hand panel.}
\end{figure}

\clearpage

\begin{figure}
\figurenum{20}
\epsscale{0.75}
\plotone{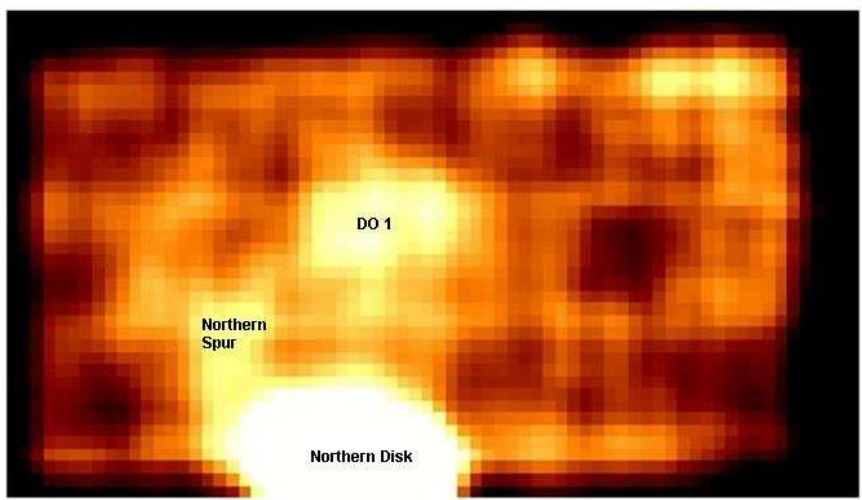}
\caption{The distribution of blue sources in the NE and NW fields. A $1 
\times 1.7$ degree$^2$ area is displayed, and 
the intensity of each pixel reflects the number of sources 
in the 100 Myr sample in $500 \times 500$ Megacam pixel ($\sim 0.5 \times 0.5$ 
kpc) gathers, after applying a low-pass filter to remove star-forming regions in 
background galaxies. Note that the Northern Spur is connected 
to the M33 disk, while a faint tongue of sources extends south of DO 1, pointing 
towards the M33 disk.}
\end{figure}

\end{document}